\documentclass[useAMS,usenatbib]{mnras}
\usepackage{graphicx}
\usepackage[english]{babel}
\usepackage{amsmath}
\usepackage{amssymb}
\usepackage{natbib}
\usepackage{wasysym}
\usepackage{newtxtext,newtxmath}
\usepackage[T1]{fontenc}
\usepackage{url}

\DeclareMathOperator{\rp}{r-process}

\title[Simulating r-process enrichment in galaxies]{Neutron star mergers and rare core-collapse supernovae as sources of r-process enrichment in simulated galaxies}
\author[F. van de Voort et al.]{Freeke~van~de~Voort,$^{1,2,3}$\thanks{E-mail: freeke@astro.cf.ac.uk} R\"udiger~Pakmor,$^1$ Robert~J.~J.~Grand,$^1$ Volker Springel,$^1$
\newauthor
Facundo~A.~G\'omez$^{4,5}$ and Federico Marinacci$^6$ \\
$^1$Max Planck Institute for Astrophysics, Karl-Schwarzschild-Stra{\ss}e 1, 85748, Garching, Germany \\
$^2$Heidelberg Institute for Theoretical Studies, Schloss-Wolfsbrunnenweg 35, 69118, Heidelberg, Germany \\
$^3$Astronomy Department, Yale University, P.O.\ Box 208101, New Haven, CT 06520-8101, USA \\
$^4$Instituto de Investigaci\'on Multidisciplinar en Ciencia y Tecnolog\'ia, Universidad de La Serena, Ra\'ul Bitr\'an 1305, La Serena, Chile \\
$^5$Departamento de F\'isica y Astronom\'ia, Universidad de La Serena, Av. Juan Cisternas 1200 Norte, La Serena, Chile \\
$^6$Department of Physics and Astronomy, University of Bologna, via Gobetti 93/2, 40129 Bologna, Italy
}

\begin{document}

\date{Accepted 2020 March 16. Received 2020 March 16; in original form 2019 July 2}

\pagerange{\pageref{firstpage}--\pageref{lastpage}} \pubyear{2020}

\maketitle

\label{firstpage}

\begin{abstract}

\noindent
We use cosmological, magnetohydrodynamical simulations of Milky Way-mass galaxies from the Auriga project to study their enrichment with rapid neutron capture (r-process) elements. We implement a variety of enrichment models from both binary neutron star mergers and rare core-collapse supernovae. We focus on the abundances of (extremely) metal-poor stars, most of which were formed during the first $\sim$Gyr of the Universe in external galaxies and later accreted onto the main galaxy.  We find that the majority of metal-poor stars are r-process enriched in all our enrichment models. Neutron star merger models result in a median r-process abundance ratio which increases with metallicity, whereas the median trend in rare core-collapse supernova models is approximately flat. The scatter in r-process abundance increases for models with longer delay times or lower rates of r-process producing events. Our results are nearly perfectly converged, in part due to the mixing of gas between mesh cells in the simulations. Additionally, different Milky Way-mass galaxies show only small variation in their respective r-process abundance ratios. Current (sparse and potentially biased) observations of metal-poor stars in the Milky Way seem to prefer rare core-collapse supernovae over neutron star mergers as the dominant source of r-process elements at low metallicity, but we discuss possible caveats to our models. Dwarf galaxies which experience a single r-process event early in their history show highly enhanced r-process abundances at low metallicity, which is seen both in observations and in our simulations. We also find that the elements produced in a single event are mixed with $\approx10^8$~M$_{\odot}$ of gas relatively quickly, distributing the r-process elements over a large region. 

\end{abstract}

\begin{keywords}
stars: abundances -- stars: neutron -- supernovae: general -- Galaxy: abundances -- galaxies: dwarf -- methods: numerical
\end{keywords}

\section{Introduction}

Observations have revealed sizeable scatter in rapid neutron capture (r-process) elemental abundances for metal-poor stars, which indicates that these elements are produced in rare objects (e.g.\ \citealt{Cowan2019}, and references therein). Binary neutron star mergers (or neutron star--black hole mergers) and special types of core-collapse supernovae, such as magneto-rotational supernovae \citep[e.g.][]{Cameron2003, Winteler2012, Mosta2018, Halevi2018} or collapsars \citep[e.g.][]{MacFadyen1999, Siegeletal2019}, are the two main classes of objects considered as r-process element production sites, which are both considered in this work. Other possible scenarios include, for example, magnetized neutrino-driven winds from proto-neutron stars \citep[e.g.][]{Thompson2018} and common envelope jet supernovae \citep[e.g.][]{Grichener2019}. 

The first detection of a neutron star merger via gravitational waves led to the observation of an optical transient \citep{Abbott2017, Coulter2017}. The colour evolution of this `kilonova' was consistent with being powered by the radioactive decay of r-process elements, thus providing strong evidence for r-process element production in neutron star mergers \citep[e.g.][]{Drout2017,Pian2017}. Although estimates of the total r-process-rich ejecta mass are high ($\approx0.05$~M$_{\astrosun}$; \citealt{Kasen2017,Cowperthwaite2017}) and the r-process nucleosynthesis robustly produces a solar abundance pattern as observed \citep[e.g.][]{Bauswein2013, Ji2016b}, it remains unclear whether or not these objects are the only or even the dominant source of r-process elements \citep[e.g.][]{Cote2019,Siegel2019}.

Idealized models of galactic chemical enrichment generally have difficulties reproducing observations of r-process abundances in low-metallicity stars via neutron star mergers \citep[e.g.][]{Argast2004, Matteucci2014, Wehmeyer2015}. However, such models do not include the full complexity of large-scale cosmic gas flows.
Powerful outflows, driven by supernovae or supermassive black holes, are frequently observed in galaxies \citep[e.g.][]{Veilleux2005, Cicone2014}. Galaxy mergers are also common, especially in the early Universe, and significantly disturb the gas and stars in a galaxy. The large-scale flow of gas throughout cosmic time is responsible for redistributing heavy elements far from their birth location and enriching both the entire interstellar medium (ISM) and the intergalactic medium. These metals can later reaccrete onto a galaxy in a small-scale or large-scale galactic fountain. The reaccretion of gas can dominate over the accretion of pristine gas, especially at low redshift (e.g.\ \citealt{Voort2017}, and references therein). Galactic outflows and galaxy mergers are therefore in part responsible for setting the elemental abundances with which subsequent generations of stars form.
Without this large-scale mixing known to occur in and around galaxies, a key aspect of galaxy formation is missing when modelling the metal abundances, which could give rise to very different results. Excluding processes that lead to more efficient mixing of heavy elements will especially have strong effects on the distribution of metals produced in rare events, such as r-process elements.

Cosmological, (magneto)hydrodynamical simulations include many of the complex processes involved in galaxy formation and can thus help study the distribution of r-process elements and their incorporation into future generations of stars. Previous simulations that included r-process enrichment through neutron star mergers have shown that including the full galaxy formation context results in a better match between simulations and observations \citep{Shen2015, Voort2015a, Naiman2018, Haynes2019}. However, these simulations included r-process enrichment in post-processing \citep{Shen2015, Naiman2018} or they did not include metal mixing \citep{Voort2015a, Haynes2019}, leading to results that were not converged \citep{Voort2015a}. Thus, large uncertainties remain in the resulting r-process abundance ratios.

The deficiencies of previous galaxy formation simulations have motivated us to revisit the question of where r-process elements are produced. Advantages of our simulation method are that we enrich the gas with r-process elements on-the-fly, therefore treating them in the same way as other elements, and that we use a moving mesh code, which allows for the exchange of material between different grid cells, resulting in more realistic mixing of metals on small scales compared to, for example, particle-based methods. Furthermore, our fiducial simulation of a Milky Way-mass galaxy down to $z=0$ has the highest resolution of any simulation that includes r-process enrichment. We explore a range of simple models for r-process enrichment by neutron star mergers and by rare core-collapse supernovae and find qualitative differences between the two sets of models. We also discuss, for the first time, differences in abundance ratios between different Milky Way-mass galaxies.

In Section~\ref{sec:sim}, we briefly describe the cosmological, magnetohydrodynamical zoom-in simulations used and give details of the chemical enrichment, including various r-process enrichment models, in Section~\ref{sec:enrich}. In Section~\ref{sec:results}, we show the abundance ratios for our different r-process enrichment models, focusing on differences between neutron star mergers (Sec.~\ref{sec:NS}) and rare core-collapse supernovae (Sec.~\ref{sec:SN}) as the sole source of r-process elements. Convergence is shown and discussed in Section~\ref{sec:res}. We compare our results to observations in Section~\ref{sec:obs}, focusing on Milky Way stars in Section~\ref{sec:MW} and an r-process enriched low-mass satellite galaxies in Section~\ref{sec:sat}, where we also explore the mixing of gas and metals. We compare our results to previous work in Section~\ref{sec:prev} and discuss and summarize our work in Section~\ref{sec:concl}.

\section{Method} \label{sec:sim}

This work uses simulations from the Auriga project\footnote{\url{https://wwwmpa.mpa-garching.mpg.de/auriga/}} \citep{Grand2017}, which consists of a large number of zoom-in magnetohydrodynamical, cosmological simulations of relatively isolated Milky Way-mass galaxies and their environments. These simulations produce realistic disc-dominated galaxies with stellar masses, galaxy sizes, rotation curves, star formation rates, and metallicities in reasonable agreement with observations.

The original suite of 30 haloes in \citet{Grand2017} has recently been extended by 10 haloes to include slightly lower halo masses \citep{Grand2019}. Additionally, these new simulations include neutron capture elements (alongside the usually included elements H, He, C, N, O, Ne, Mg, Si, Fe) as passive tracers (see Section \ref{sec:enrich}). Of the original suite, 6 galaxies have been rerun with r-process elements included, which means that we can study 16 different Milky Way-mass galaxies in total. One of the new haloes, named halo L8, has been run down to $z=0$ at very high resolution. We briefly describe the simulations below, focusing on the high-resolution simulation, from which we derive our main results. More details can be found in \citet{Grand2017} and references therein. 

The simulations were carried out with the quasi-Lagrangian moving mesh code \textsc{arepo} \citep{Springel2010, Pakmor2016, Weinberger2019} and assume a $\Lambda$CDM cosmology with parameters taken from \citet{PlanckXVI2014}: $\Omega_\mathrm{m}= 1-\Omega_\Lambda = 0.307$, $\Omega_\mathrm{b}= 0.048$, $h = 0.6777$, $\sigma_8 = 0.8288$, and $n = 0.9611$. 

Here, we focus on one high-resolution zoom-in simulation, for which the target cell resolution is $8.1\times10^3$~M$_{\astrosun}$ and dark matter particle mass is $4.3\times10^4$~M$_{\astrosun}$.\footnote{This resolution is also referred to in the literature as `level 3'.} The gravitational softening length of the dark matter and star particles and the minimum softening length of the gas is 184~proper~pc at $z<1$ and 369~comoving~pc at higher redshift. The simulation data was saved at 128 discrete output redshifts. 
For our resolution test, we use simulations with 8 and 64 times lower mass resolution, which we refer to as medium-resolution and low-resolution, respectively.\footnote{Simulations with 8 and 64 times lower mass resolution than our fiducial simulation are also referred to as `level 4' and `level 5', respectively.} To quantify the differences between Milky Way-mass galaxies, we use our suite of 16 medium-resolution simulations. 

Gas becomes star-forming above a density threshold of $n_\mathrm{H}^\star=0.11$~cm$^{-3}$ and forms stars stochastically \citep{Springel2003}. In general, the entire gas cell is converted to a star particle. The exception is the rare case that the total mass exceeds the target mass resolution by more than a factor of two, in which case only the target mass is converted to a star and the gas cell is retained with reduced mass.

The model includes both stellar and active galactic nucleus (AGN) feedback, which results in large-scale outflows (see \citealt{Grand2017} for details). These outflows push metals far out into the halo, which can later reaccrete onto the galaxy and get incorporated into new generations of stars. They therefore have a large impact on the chemical evolution of the galaxy. 

\begin{figure}
\center
\includegraphics[scale=1]{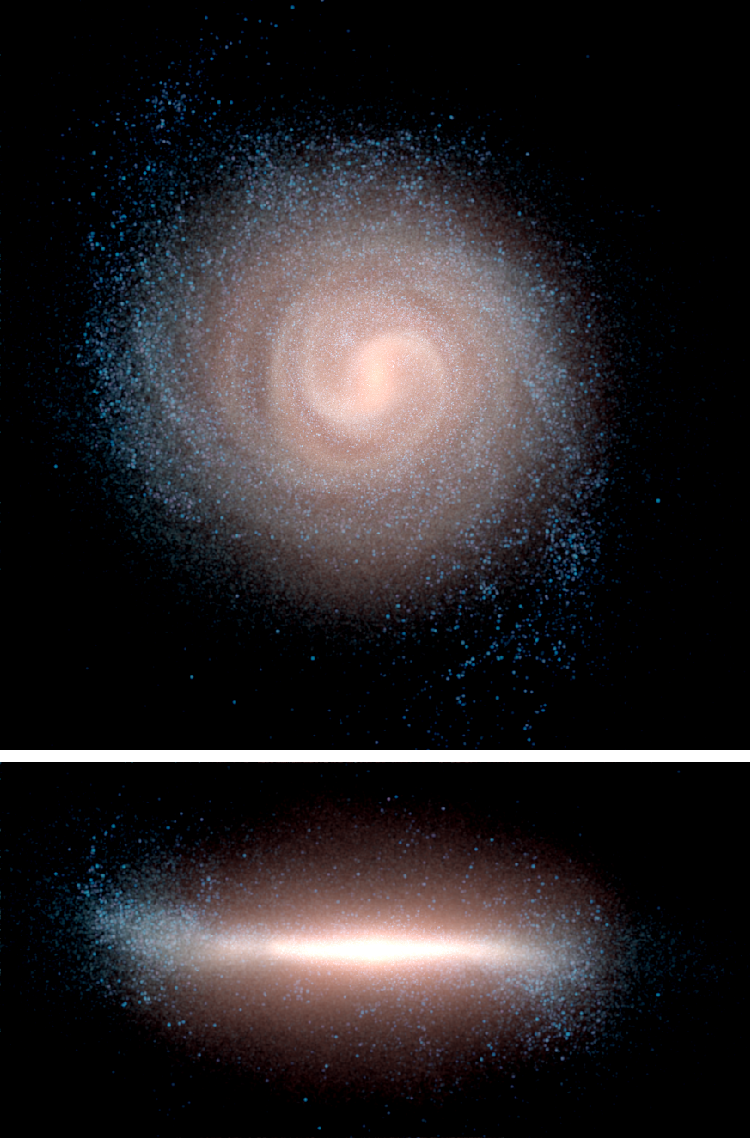}
\caption {\label{fig:img} 50 kpc $\times$ 50 kpc face-on and 50 kpc $\times$ 25 kpc edge-on images at $z=0$ in K-, B-, and U-band, which are shown by red, green and blue colour channels, respectively. Younger stars thus appear bluer. The galaxy consists both of a star-forming disc, which dominates in the outskirts, and an older bulge, which dominates in the centre.}
\end{figure}

The resulting galaxy at $z=0$ has a total stellar mass of $10^{10.7}$~M$_{\astrosun}$, a total halo mass of $10^{11.9}$~M$_{\astrosun}$, and a virial radius, $R_\mathrm{vir}$ of 200~kpc. A face-on and an edge-on three-colour image (in K-, B-, and U-band) are shown in Figure~\ref{fig:img}. The galaxy has both a clear disc and a prominent bulge. The outer part of the galaxy is dominated by young, blue stars, whereas the inner part is dominated by older, redder stars. 

\begin{figure}
\center
\includegraphics[scale=.4]{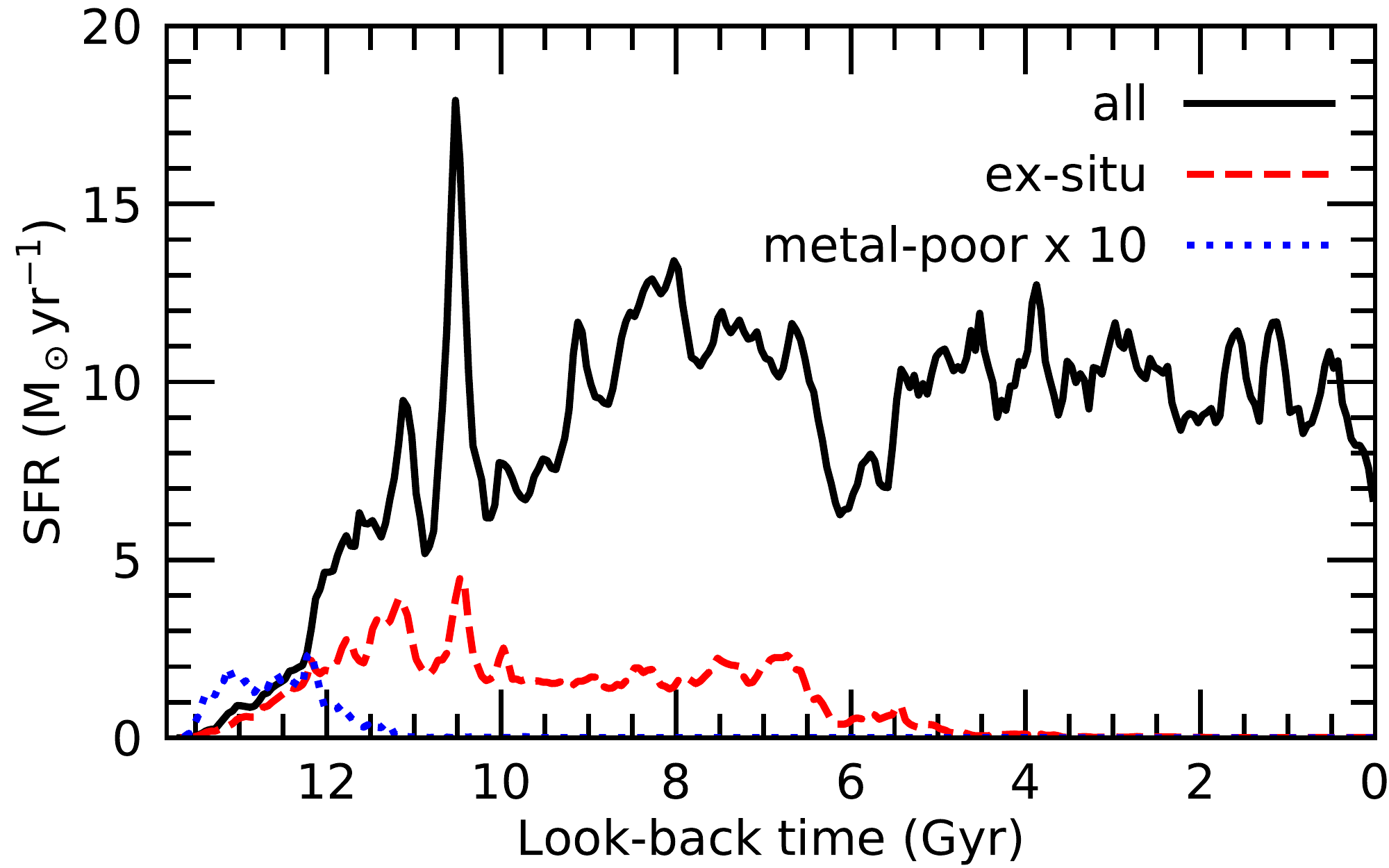}
\caption {\label{fig:sfh} Star formation history (using 50 Myr bins) based on ages and initial masses of all the star particles within $R_\mathrm{vir}$ at $z=0$ (black, solid curve). The red, dashed curve shows the star formation history of the subset of stars that was formed outside the main progenitor of the central Milky Way-mass galaxy (i.e.\ ex-situ). The blue, dotted curve shows the star formation history of the subset of stars with $\mathrm{[Fe/H]}<-2$ multiplied by a factor of 10. Ex-situ stars dominate at early times, when there are numerous small building block galaxies. At intermediate times, most stars are formed in-situ, but there is still a non-negligible contribution of ex-situ stars. At late times this galaxy does not experience any substantial mergers and the ex-situ contribution vanishes. Metal-poor stars form exclusively at very high redshift and their contribution becomes negligible when the Universe is less than 3~Gyr old.}
\end{figure}
The star formation history of our high-resolution galaxy is shown by the black, solid curve in Figure~\ref{fig:sfh}, calculated by adding up the initial masses of all stars within $R_\mathrm{vir}$ at $z=0$ in 50~Myr age bins. The red, dashed curve shows the subset of stars formed outside the main progenitor of the central galaxy, either in satellite galaxies or in galaxies outside the virial radius of the main galaxy, identified in the same way as in \citet{Monachesi2019}. Only 12 per cent of the total stellar mass within $R_\mathrm{vir}$ has been accreted rather than formed in-situ. In the first $\sim$1.6~Gyr ($z>4$), however, most of the stars were formed external to the central galaxy. This is also the time at which the majority of metal-poor stars form. The blue, dotted curve shows the star formation history of all metal-poor stars, with $\mathrm{[Fe/H]}<-2$, multiplied by a factor of 10 for visibility. Metal-poor stars are very old in our simulation; the very last metal-poor star particle forms at $z=1.7$ or 10~Gyr ago, but their median age is 12.6~Gyr. 

The low redshift star formation rate fluctuates around $\approx10$~M$_{\astrosun}$yr$^{-1}$, which is higher by a factor of a few than the one inferred for the Milky Way \citep[e.g.][]{Diehl2006,Snaith2015}. This will also affect the late-time chemical enrichment, so an exact match between our simulation and the Milky Way is not expected. However, firstly, we focus our attention on metal-poor stars, for which the late-time star formation rate is unimportant. And secondly, we discuss the variation between different galaxy simulations in Section~\ref{sec:diff} and show that the differences are relatively minor.

\subsection{Chemical enrichment} \label{sec:enrich}

Every star particle in the simulation represents a single stellar population with a specific age and metallicity. We assume an initial mass function from \citet{Chabrier2003}. Stars with masses between 8~M$_{\astrosun}$ and 100~M$_{\astrosun}$ are assumed to explode as core-collapse supernovae. The mass-loss and metal return is calculated for core-collapse and Type Ia supernovae and for asymptotic giant branch (AGB) stars at each simulation time-step using the yields from \citet{Karakas2010} for AGB stars, from \citet{Portinari1998} for core-collapse supernovae, and from \citet{Thielemann2003} and \citet{Travaglio2004} for Type Ia supernovae. The mass and metals are then equally distributed amongst approximately 64 neighbouring gas cells.\footnote{To study the effect of this choice, we ran an additional simulation in which the material is only added to the single host cell of the star particle. The results are shown and discussed in Section~\ref{sec:res}.}

In our simulation, we trace some of the more common elements (H, He, C, N, O, Ne, Mg, Si, Fe) which allow the gas to cool radiatively and thus affect the growth of the galaxy. We additionally implement a range of models for r-process enrichment, which we also trace on-the-fly, i.e.\ with the same time resolution as any other property in the simulation. However, the newly added r-process elements are implemented as passive tracers and therefore do not affect the dynamical evolution of the galaxy. Given how rare these elements are, they are not expected to dynamically influence galaxy formation in reality either. Whereas other elements are produced using yield tables, we assume that each r-process event produces the same amount of r-process elements. The effective yield is set to a different value for each r-process enrichment model in post-processing. This is done by renormalizing each model independently to the same value at a specific metallicity (see below). 

The r-process enrichment events, in this work either through neutron star mergers or through a subclass of core-collapse supernovae, are implemented stochastically. This means that we pick random numbers to determine whether a star particle has an r-process production event at this specific simulation time-step. The r-process enrichment is then only done at the time-step at which such an event took place. Because a star particle represents a simple stellar population (with initial mass $\approx8\times10^3$~M$_{\astrosun}$), it is in principle allowed to produce multiple r-process enrichment events over its life time, although this is rare. None of the r-process enrichment models in this work include a dependence on metallicity, although it is possible that the formation and/or evolution of binary neutron stars and/or rapidly rotating massive stars changes with metallicity. However, the exact form of the metallicity dependence is very uncertain and we therefore chose not to include it here. 

Other metal species, such as iron and magnesium, are produced in much more common sources. The enrichment from standard core-collapse and Type Ia supernovae is not sampled stochastically. Rather, each stellar particle at every time-step smoothly enriches its surroundings with elements up to the iron peak based on its age. This choice might slightly reduce the scatter in the abundance of these metal species, but is not expected to significantly impact our results given that the production sites of these metals are ubiquitous.

In our simulations, no iron is produced by the sources that produce r-process elements. This is a good approximation for neutron star mergers, which produce very little iron. However, if the subclass of core-collapse supernovae that we consider produce a substantial amount of iron, which is mixed with the r-process elements produced in the same event, this would increase the iron abundance of stars formed from the gas enriched by such an event. This is discussed further in Section~\ref{sec:SN}.

\subsubsection{r-process models: neutron star mergers}

\begin{table}
\begin{center}                                                                                                                                        
\caption{\label{tab:NSmodels} \small Parameters of neutron star merger models (see also Eq.~\ref{eqn:rate}): model name, number of neutron star mergers per M$_{\astrosun}$ of stars, minimum delay time for the first neutron star merger in a simple stellar population, delay time distribution power-law exponent, resulting neutron star merger rate at $z=0$ (averaged over the last Gyr), and the europium yield, $y_\mathrm{Eu}$, per neutron star merger (based on our chosen normalization, see Section~\ref{sec:results}). The total yield of r-process material for elements with $A>92$ is 215 times higher than $y_\mathrm{Eu}$. For each model we vary one parameter compared to the fiducial model, which is indicated in bold.}         
\begin{tabular}[t]{llrrrr}
\hline \\[-3mm]                                                                                                                                       
model name   & $A$                                   & $t_\mathrm{min}$              & $\gamma$           & $R_\mathrm{NS}(z=0)$ & $y_\mathrm{Eu}$ \\
                      & (M$_{\astrosun}^{-1}$)                  & (Myr)                                  &                           & (yr$^{-1}$)             & (M$_{\astrosun}$)\\
\hline \\[-4mm]                                                                                                                                       
  fiducial NS     & $3\times10^{-6}$                & $30$               & $-1.0$               & $1.7\times10^{-4}$       & $4.7\times10^{-5}$ \\
  high rate      & $\mathbf{1\times10^{-5}}$ & $30$               & $-1.0$               & $5.8\times10^{-4}$       & $1.4\times10^{-5}$ \\
  low rate        & $\mathbf{1\times10^{-6}}$ & $30$               & $-1.0$               & $5.8\times10^{-5}$      & $1.4\times10^{-4}$ \\
  short delay   & $3\times10^{-6}$                & $\mathbf{10}$ & $-1.0$              & $2.0\times10^{-4}$      & $3.8\times10^{-5}$ \\
  long delay    & $3\times10^{-6}$                & $\mathbf{100}$ & $-1.0$              & $1.4\times10^{-4}$      & $6.4\times10^{-5}$ \\
  shallow DTD & $3\times10^{-6}$                & $30$               & $\mathbf{-0.5}$ & $5.5\times10^{-4}$    & $2.2\times10^{-5}$ \\
  steep DTD    & $3\times10^{-6}$                & $30$               & $\mathbf{-1.5}$ & $2.7\times10^{-5}$     & $2.5\times10^{-4}$ \\
\hline                                                                                                                                                
\end{tabular}                                                                                                                                         
\end{center}                                                                                                                                          
\end{table}      

Neutron star mergers are assumed to be the source of all r-process enrichment in 7 of our 10 models. Because the rates of neutron star mergers and their delay time distribution (DTD) are not well-constrained, we use a range of models that aim to bracket the possibilities. Each star particle in our simulation represents a simple stellar population and we parametrize the
rate of NS mergers ($R_\mathrm{NS}$) in each one as
\begin{equation} \label{eqn:rate}
R_\mathrm{NS} = A M_\mathrm{\star} t^\gamma \ \mathrm{for} \ t > t_\mathrm{min}
\end{equation}
and $R_\mathrm{NS}=0$ otherwise, where $A$ is the number of neutron star mergers per unit of stellar mass, $M_\mathrm{\star}$ is the mass of the star particle, $t$ is the time since the formation of the star particle, $\gamma$ is the (always negative) exponent of the time dependence of the delay time distribution, and $t_\mathrm{min}$ is the minimum time needed for a neutron star merger to take place. Each star particle has a non-zero, but small, chance to experience a neutron star merger and enrich their surroundings with r-process elements, based on stochastically sampling the delay time distribution. 

The present-day merger rate is estimated to be $\approx10^{-4}$~yr$^{-1}$ in the Milky Way, but could be as low as $\approx10^{-6}$~yr$^{-1}$ or as high as $\approx10^{-3}$~yr$^{-1}$ \citep[e.g.][]{Abadie2010}. The delay time distribution is expected to be inversely proportional to the age of the stellar population, after a minimum delay of about $10^7$~yr \citep[e.g.][]{Belczynski2006}.
Therefore, we take the following values for our fiducial neutron star merger model: $A=3\times10^{-6}$~M$_\odot^{-1}$, $\gamma=-1$, and $t_\mathrm{min}=3\times10^7$~yr. This results in $R_\mathrm{NS}=1.7\times10^{-4}$~yr$^{-1}$ at $z=0$ for our high-resolution simulation.  

Because the parameter values of the delay time distribution have large uncertainties associated with them, we vary $A$, $\gamma$, and $t_\mathrm{min}$ as listed in Table~\ref{tab:NSmodels}. The second to last column gives the resulting neutron star merger rate for our simulated galaxy at $z=0$. The resulting rates differ by more than an order of magnitude, but all of the models are consistent with constraints from observations within uncertainties \citep[e.g.][]{Behroozi2014,Kim2015,Abbott2017}. The last column lists the amount of europium, an element which is almost exclusively produced by the r-process, ejected by a single r-process event. These yields are set for each neutron star merger model separately so that the average europium abundance at solar metallicity matches the solar value (i.e.\ $\mathrm{[Eu/Mg]}=0$ at $\mathrm{[Mg/H]}=0$) as given by \citet{Asplund2009}. To calculate the total amount of r-process material above the first neutron capture peak (i.e. elements more massive than zirconium or atomic mass number $A>92$), the europium yields, $y_\mathrm{Eu}$, should be multiplied by a factor~215. This factor is based on the r-process abundances of each element taken from \citet{Burris2000}. All of our neutron star merger yields are consistent with the lower limit derived by \citet{Macias2018}.

\subsubsection{r-process models: rare supernovae} 

\begin{table}
\begin{center}                                                                                                                                        
\caption{\label{tab:SNmodels} \small Parameters of rare supernova models: model name, the fraction of core-collapse supernovae producing r-process elements, the resulting rate of r-process producing core-collapse supernovae in the last Gyr, and the europium yield per r-process producing event (based on our chosen normalization, see Section~\ref{sec:results}). Multiplying $y_\mathrm{Eu}$ by 215 gives the total yield of r-process material for elements with $A>92$.}         
\begin{tabular}[t]{llrr}
\hline \\[-3mm]                                                                                                                                       
model name         & fraction & $R_\mathrm{ccSNrp}(z=0)$   &  $y_\mathrm{Eu}$ \\
                            &              & (yr$^{-1}$)                            &(M$_{\astrosun}$)\\
\hline \\[-4mm]                                                                                                                                       
  high fraction       & 10\%  & $1.1\times10^{-2}$                & $5.4\times10^{-7}$ \\
  medium fraction & 1\%    & $1.1\times10^{-3}$                 & $5.4\times10^{-6}$ \\
  low fraction        & 0.1\% (fiducial) & $1.1\times10^{-4}$  & $5.5\times10^{-5}$ \\
\hline                                                                                                                                                
\end{tabular}                                                                                                                                         
\end{center}                                                                                                                                          
\end{table}      

Besides the neutron star merger models, we implemented 3 additional models in which the r-process element production rate scales with the rate of core-collapse supernovae.\footnote{All our simulations have these models, including our fiducial, high-resolution simulation. The 16 additional medium-resolution simulations include a fourth core-collapse supernova model with an even lower event rate, discussed in Section~\ref{sec:diff}.} Possible sources include magneto-rotational supernovae and collapsars (see \citealt{Cote2019} for an overview). We remain agnostic towards the exact type of object producing r-process elements and just use a fixed fraction of the total core-collapse supernova rate, between 0.1 and 10 per cent as listed in Table~\ref{tab:SNmodels}, which we sample stochastically, as was also done for the neutron star merger models (see Section~\ref{sec:enrich}). The rate of magneto-rotational supernovae used in \citet{Haynes2019} at low metallicity is similar to (about a factor~2 lower than) the rate in our fiducial model (``low fraction'') at low metallicity, but is substantially lower at high metallicity. A possible dependence on other properties, for example on stellar metallicity, is not taken into account, because we opted to use only the simplest models here. The third column in Table~\ref{tab:SNmodels} lists the r-process-producing core-collapse supernova rate for the simulated galaxy at $z=0$. The last column provides the amount of europium produced by a single r-process event. Multiplying $y_\mathrm{Eu}$ by a factor~215 gives the total amount of r-process material above the first neutron capture peak (i.e. elements more massive than zirconium). The r-process yield in the ``high fraction'' model is inconsistent with the lower limit derived by \citet{Macias2018}, based on the observed abundance ratios of a single star, but our other models have yields above this lower limit. 

Both the star formation rate and the rate of core-collapse supernovae at $z=0$ (i.e.\ about 10 per century) in our simulation are likely somewhat higher than those in the Milky Way, although large uncertainties remain in determining the latter \citep{Chomiuk2011, Adams2013}. Because we focus primarily on the low-metallicity stars, which all formed at high redshift, we do not expect this difference to substantially influence our conclusions.

\begin{figure}
\center
\includegraphics[scale=.4]{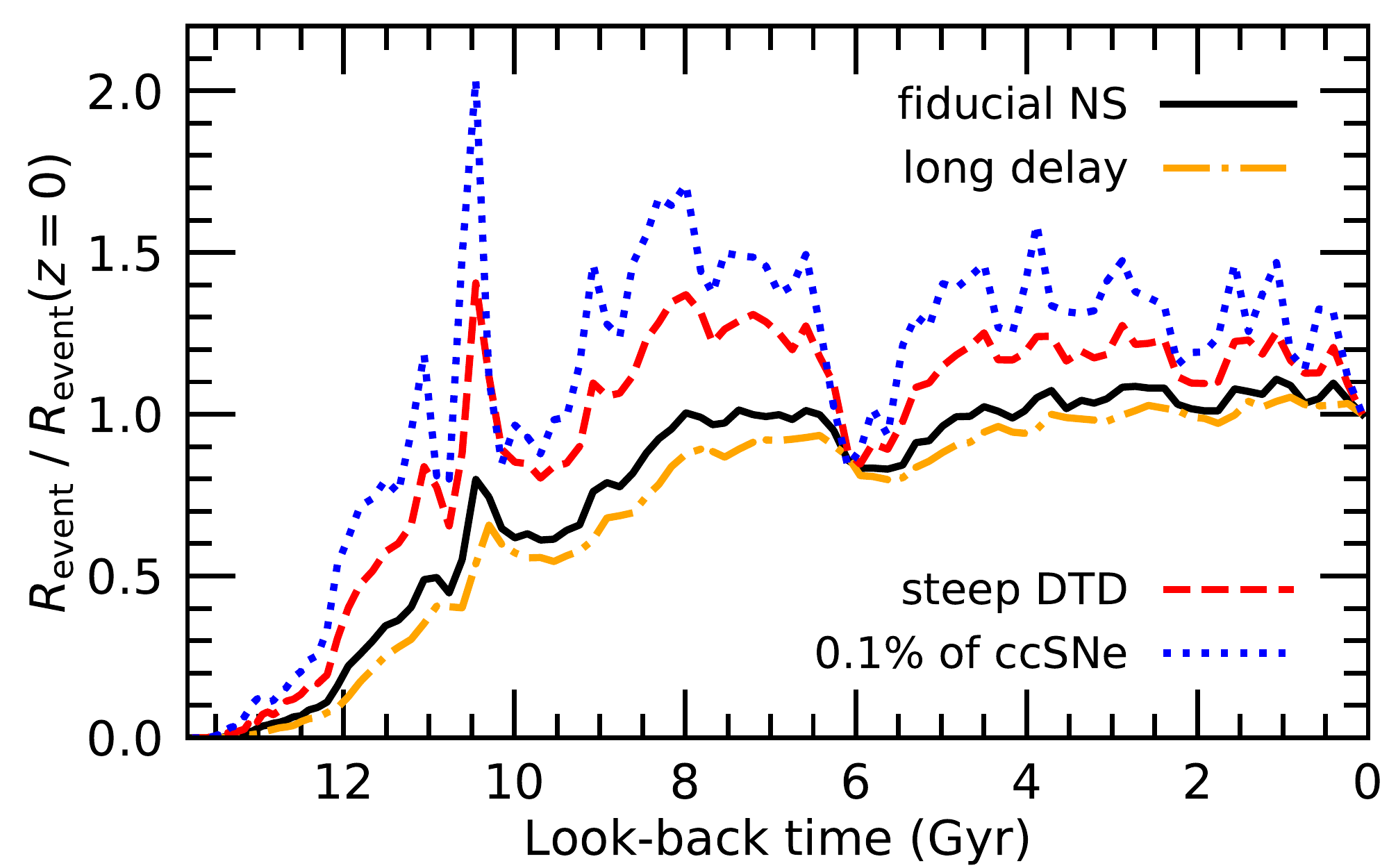}
\caption {\label{fig:rph} The r-process event rate history (using the 128 simulation outputs) for all the star particles within $R_\mathrm{vir}$ at $z=0$ in 4 out of our 10 models divided by the event rate at $z=0$. The models with rare core-collapse supernova (blue, dotted curve) and steep delay time distribution (red, dashed curve) have relatively more events at early times than the fiducial model (black, solid curve), whereas the model with longer delay times (orange, dot-dashed curve) has fewer early events. These differences impact the r-process abundance ratios as discussed in Section~\ref{sec:results}.}
\end{figure}

The resulting history of r-process producing events is shown in Figure~\ref{fig:rph}, normalized by the $z=0$ event rate, for the fiducial neutron star merger model (black, solid curve) as well as models ``long delay'' (orange, dot-dashed curve) and ``steep DTD'' (red, dashed curve). Our fiducial rare core-collapse supernova model (``low fraction'') is shown as the blue, dotted curve. The rare core-collapse supernova model follows the star formation history shown in Figure~\ref{fig:sfh} closely, as expected. The fiducial neutron star merger model clearly depends on the global shape of the star formation history, but is much less affected by its fluctuations. Furthermore, this model has relatively fewer events at early times. Model ``steep DTD'' has a normalized r-process event rate in between the fiducial neutron star merger model and the rare core-collapse supernova models. The model with longer minimum delay time has relatively fewer r-process events at early times compared to the fiducial model, although the difference is small. The difference in normalized event rates decreases towards late times for all models. Note that the normalization of these models is also very different, as detailed in Tables~\ref{tab:NSmodels} and~\ref{tab:SNmodels}. Although the other models are not shown, they behave as expected, i.e.\ the normalized event rate in model ``short delay'' lies slightly above model ``fiducial NS'', whereas model ``shallow DTD'' lies substantially below.

\section{Results} \label{sec:results}

\begin{figure*}
\center
\includegraphics[scale=.45]{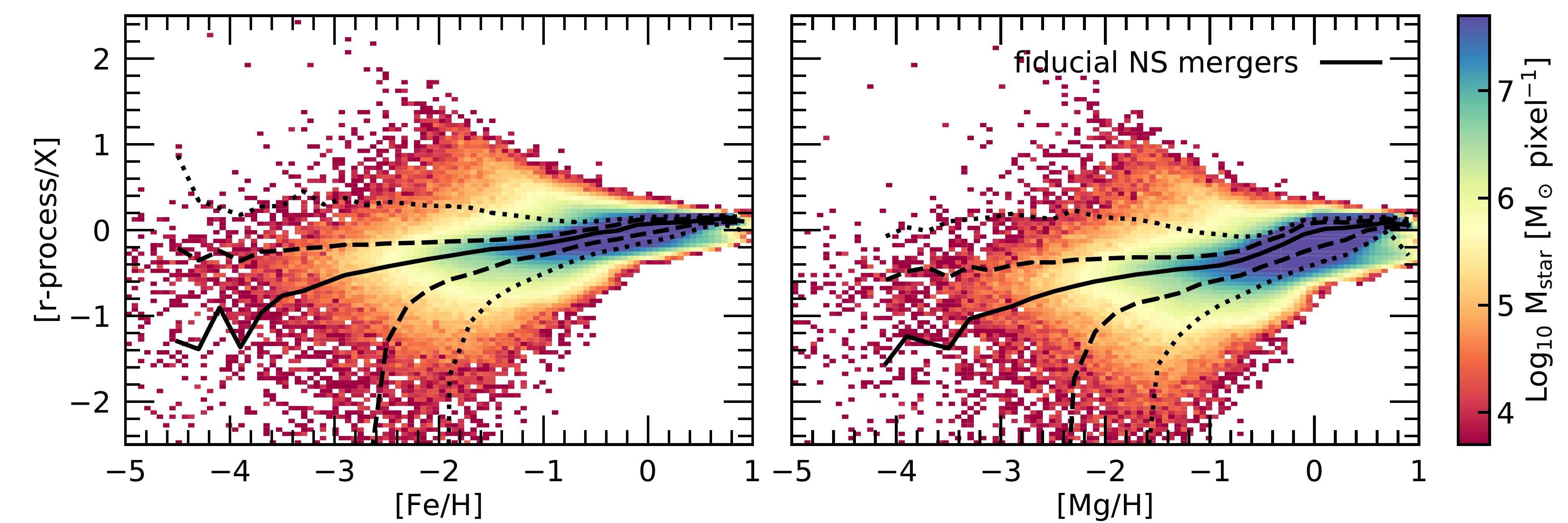}
\caption {\label{fig:NS} $\mathrm{[\rp/Fe]}$ ($\mathrm{[\rp/Mg]}$) as a function of $\mathrm{[Fe/H]}$ ($\mathrm{[Mg/H]}$) are shown in the left-hand (right-hand) panel for our fiducial neutron star merger enrichment model. The colour coding of the 200 by 200 pixel images represents the logarithm of the stellar mass per pixel. Black curves show the median (solid curves), the $1\sigma$ scatter (dashed curves), and the $2\sigma$ scatter (dotted curves). The r-process abundance ratios have been normalized so that the median $\mathrm{[\rp/Mg]}$ is zero at $\mathrm{[Mg/H]}=0$ (see Table~\ref{tab:NSmodels}). The median $\mathrm{[\rp/Fe]}$ and $\mathrm{[\rp/Mg]}$ increase with increasing metallicity. The distribution shows there are many extreme outliers in the metal-poor regime (more than expected from a Gaussian distribution) and the scatter increases towards low metallicity.}
\end{figure*}

In the following, we show results for both disc stars and halo stars out to the virial radius of our simulated Milky Way-mass galaxy (or galaxies) at $z=0$. 
Abundance ratios of a star compared to those of the Sun are defined as
\begin{equation}
[A/B]=\mathrm{log}_{10}\left(\frac{N_A}{N_B}\right)_\mathrm{star}-\mathrm{log}_{10}\left(\frac{N_A}{N_B}\right)_{\astrosun}
\end{equation}
where $A$ and $B$ are different elements and $N_{\mathrm{A}}$ and $N_{\mathrm{B}}$ are their number densities. Solar abundances are taken from \citet{Asplund2009}. Throughout this work, we refer to stars with $\mathrm{[Fe/H]}<-2$ as metal-poor stars and to those with $\mathrm{[Fe/H]}<-3$ as extremely metal-poor stars.

Type Ia supernovae contribute only about 40 per cent of the total amount of iron in our simulations by $z=0$, which is somewhat lower than their observed contribution in the Milky Way. It is not known what exactly causes this underproduction of Type Ia iron or overproduction of iron from core-collapse supernovae in our simulations. Nevertheless, the result of this, in combination with our chosen yield tables, is that there is not a large enough of a decrease in $\mathrm{[\alpha/Fe]}$ at high metallicity as compared to observations of Milky Way stars: $\mathrm{[O/Fe]}$ ($\mathrm{[Mg/Fe]}$) decreases by 0.08 dex (0.17~dex) from $\mathrm{[Fe/H]}=-1$ to $\mathrm{[Fe/H]}=0$ and $\mathrm{[Si/Fe]}$ even increases slightly by 0.08 dex in our fiducial simulation. Observations show a much larger decrease for all three $\alpha$ element abundance ratios. We therefore, in this work, focus on the low-metallicity end ($\mathrm{[Fe/H]}<-1$) and note here that the high-metallicity regime is expected to change in future simulations with more realistic late-time iron production. We also show a subset of our results with magnesium instead of iron as a metallicity indicator, which does not change any of our conclusions. Because the magnesium yields used in our simulations are known to be low, we multiply them by a factor~2.5 in post-processing to correct for this. This is further detailed in Appendix~\ref{sec:FeMg}, which shows the magnesium abundance ratios in our fiducial simulation. $\alpha$ abundances in the Auriga simulations are discussed in further detail in \cite{Grand2018}.

Because the r-process yields are uncertain, the r-process abundance ratios shown in this work are renormalized. As stated above, our simulations underproduce the enhanced late-time iron production and thus do not show the $\approx0.4$~dex observed decrease in $\mathrm{[\rp/Fe]}$ at the high-metallicity end. We therefore use the magnesium abundances to normalize to the solar value, i.e.\ $\mathrm{[\rp/Mg]}=0$ at $\mathrm{[Mg/H]}=0$ and use the same r-process yields to calculate $\mathrm{[\rp/Fe]}$. Models with a larger number of r-process producing sources will therefore have a lower r-process yield per event. This can be done self-consistently in post-processing because the r-process elements are passive tracers in the simulation (as opposed to other metals species which affect the cooling of the gas). In the following we focus on the shape of the relation between stellar abundances rather than its normalization. To calculate the median and scatter, we use 0.2~dex bins and only show those bins that contain at least 100 star particles.

\subsection{Neutron star mergers} \label{sec:NS}


Figure~\ref{fig:NS} shows the abundance ratio of r-process elements to iron (magnesium) for our fiducial neutron star merger enrichment model as a function of iron (magnesium) metallicity in the left-hand (right-hand) panel. The median abundance ratios are shown by the black solid curves and the $1\sigma$ and $2\sigma$ scatter by the dashed and dotted curves, respectively. The median increases with metallicity for both iron and magnesium, while the scatter decreases. Low-metallicity stars are rare and only 0.03 per cent of the star particles have $\mathrm{[Fe/H]}<-3$, which is the reason we need a high-resolution simulation to be able to study extremely metal-poor stars. It can be useful to show both r-process elements with respect to iron and magnesium, because of the uncertainties involved with the yields of any element. In most of the following, we use iron as a metallicity indicator, but the results presented in this work are independent of this choice.

Even at the lowest metallicities we can probe with our simulation, the majority of stars are enriched with r-process elements. However, the median abundance ratio is lower than at intermediate metallicities, giving rise to an increasing trend of $\mathrm{[\rp/Fe]}$ and $\mathrm{[\rp/Mg]}$ with metallicity. The $2\sigma$ scatter is larger than twice the $1\sigma$ scatter, which means that there are more extreme outliers than expected from a Gaussian distribution.

\begin{figure}
\center
\includegraphics[scale=.4]{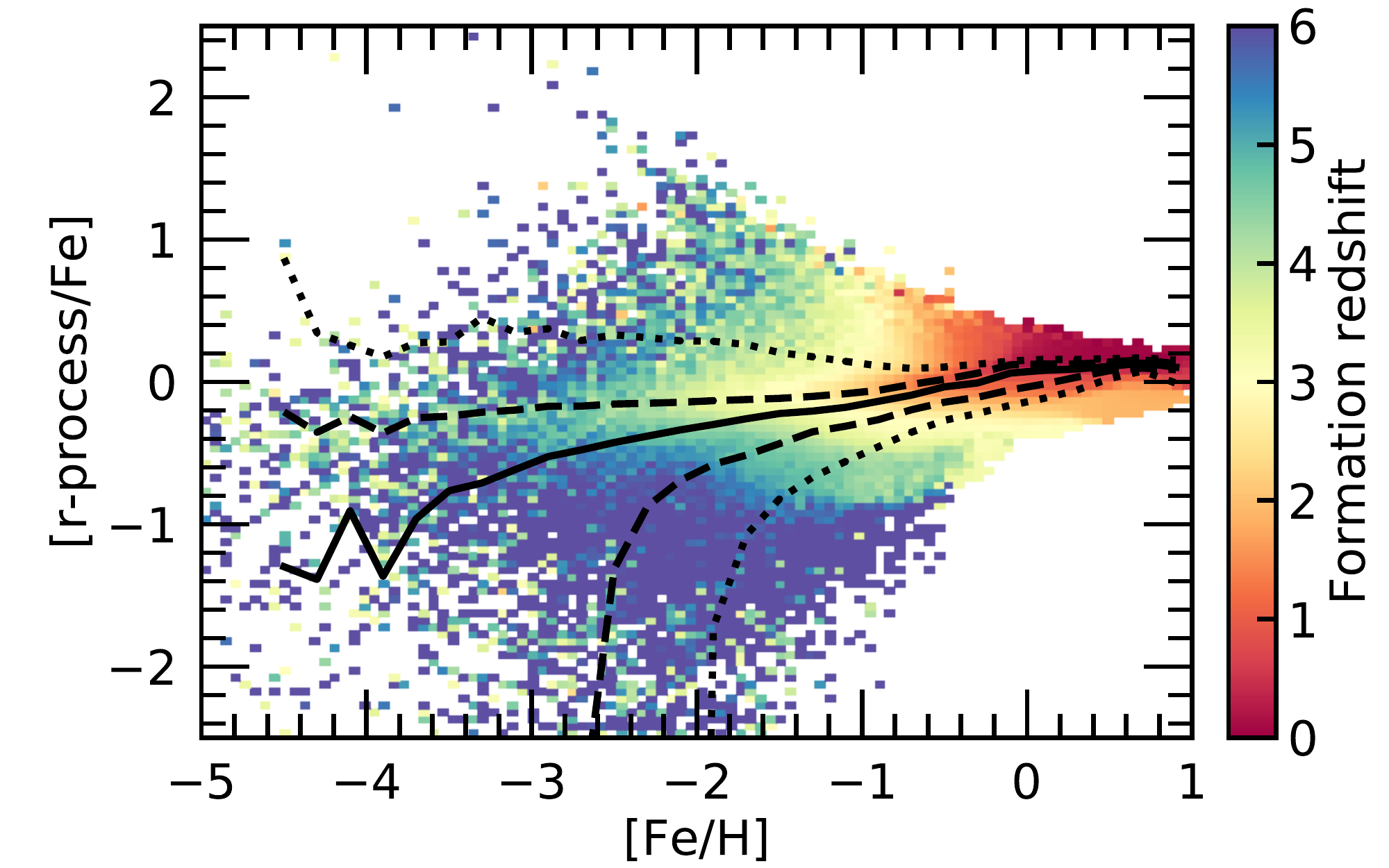}
\caption {\label{fig:NSz} $\mathrm{[\rp/Fe]}$ as a function of $\mathrm{[Fe/H]}$ for our fiducial neutron star merger enrichment model colour coded by the (mass-weighted) formation redshift of the stars. The r-process yield is chosen so that $\mathrm{[\rp/Mg]}=0$ at $\mathrm{[Mg/H]}=0$ (see Figure~\ref{fig:NS} and Table~\ref{tab:NSmodels}). A large fraction of very low-metallicity stars were formed when the Universe was less than a Gyr old ($z>6$). Outliers from the median trend were also on average formed earlier than the bulk of the stars at fixed metallicity.}
\end{figure}

Metal-poor stars (i.e.\ $\mathrm{[Fe/H]}<-2$) in our simulation have a median formation redshift of $z=4.7$, which corresponds to a median age of 12.6~Gyr, as was shown previously in Figure~\ref{fig:sfh}. Extremely metal-poor stars (i.e.\ $\mathrm{[Fe/H]}<-3$) on average formed at an even higher redshift, $z=6.2$, i.e.\ 12.9~Gyr ago, when the Universe was less than a Gyr old. Besides the abundance of all metal species increasing with time, the r-process abundance ratio also exhibits a residual dependence on formation redshift at fixed metallicity. This is shown in Figure~\ref{fig:NSz}, which is the same as the left-hand panel of Figure~\ref{fig:NS} but coloured according to the average formation redshift of the stars in each pixel. Stars that are outliers in $\mathrm{[\rp/Fe]}$ were formed earlier than stars that are near the median trend at fixed [Fe/H]. Our other models (with either neutron star mergers or rare core-collapse supernovae) exhibit the same behaviour. 

Qualitatively similar results were obtained in simulations which employed a very different hydrodynamical method with no metal exchange between different resolution elements \citep{Voort2015a}. This trend therefore seems to be robust to the choice of numerical method. Quantitatively, however, there are some noticeable differences. The scatter in $\mathrm{[\rp/Fe]}$ was larger in \citet{Voort2015a} and the median formation redshift (age) of extremely metal-poor stars was $z\approx2.9$ (11.6~Gyr) instead of $z=6.2$ (12.9~Gyr). Both of these differences are likely due to enhanced small-scale mixing in our new calculations which allows material to move from one gas cell to another (see also Section~\ref{sec:sat}). Improved mixing also results in fewer extremely metal-poor stars, because heavy elements produced are more quickly distributed over the entire ISM.

\begin{figure}
\center
\includegraphics[scale=.4]{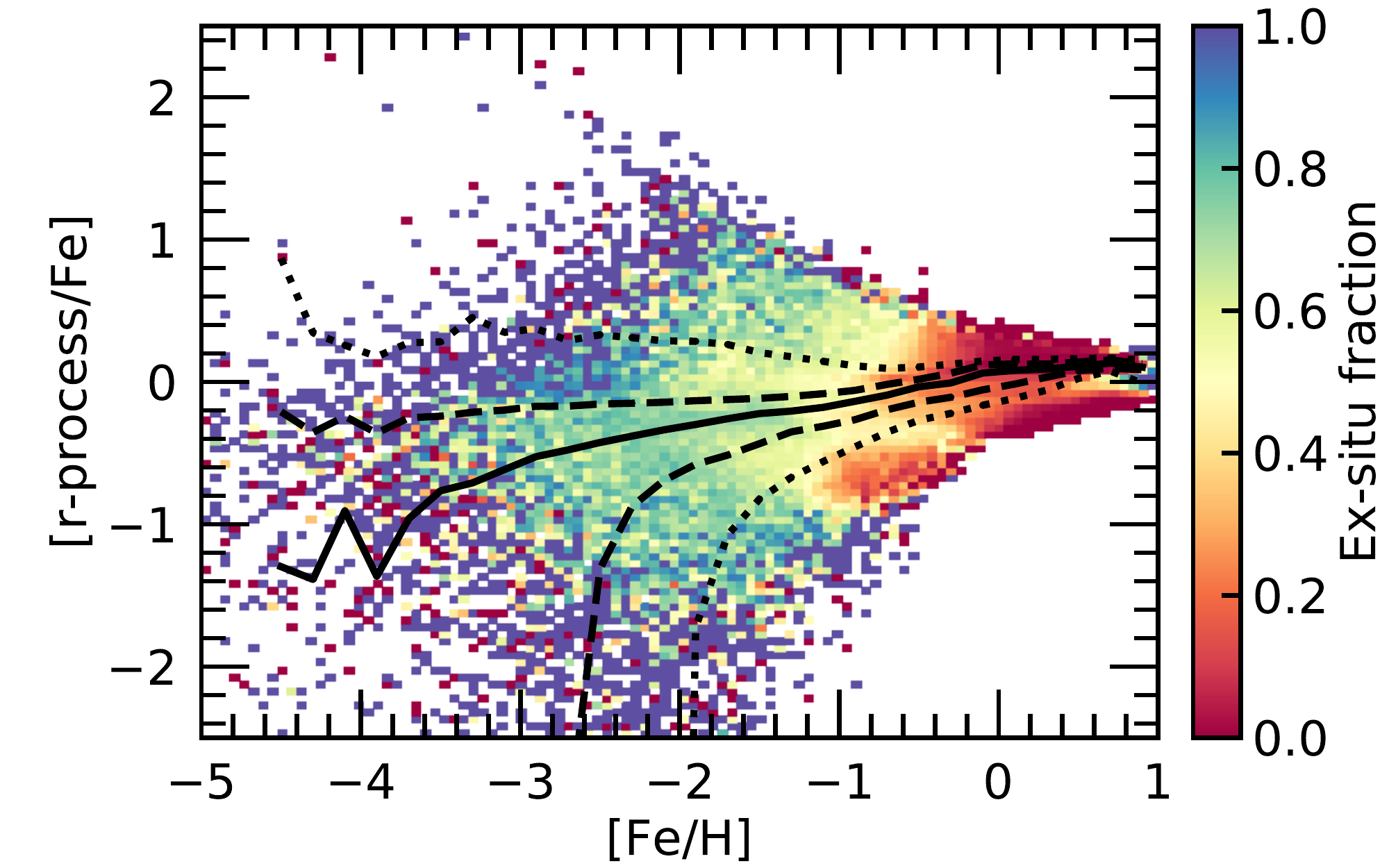}
\caption {\label{fig:NSex} $\mathrm{[\rp/Fe]}$ as a function of $\mathrm{[Fe/H]}$ for our fiducial neutron star merger enrichment model colour coded by the (mass-weighted) ex-situ fraction of the stars. The r-process yield is chosen so that $\mathrm{[\rp/Mg]}=0$ at $\mathrm{[Mg/H]}=0$ (see Figure~\ref{fig:NS} and Table~\ref{tab:NSmodels}). A high fraction of extremely metal-poor stars were formed in other galaxies than the main progenitor of our simulated galaxy. At low metallicity, outliers in $\mathrm{[\rp/Fe]}$ also have a further enhanced ex-situ fraction compared to stars with r-process abundances closer to the median.}
\end{figure}
\begin{figure*}
\center
\includegraphics[scale=.5]{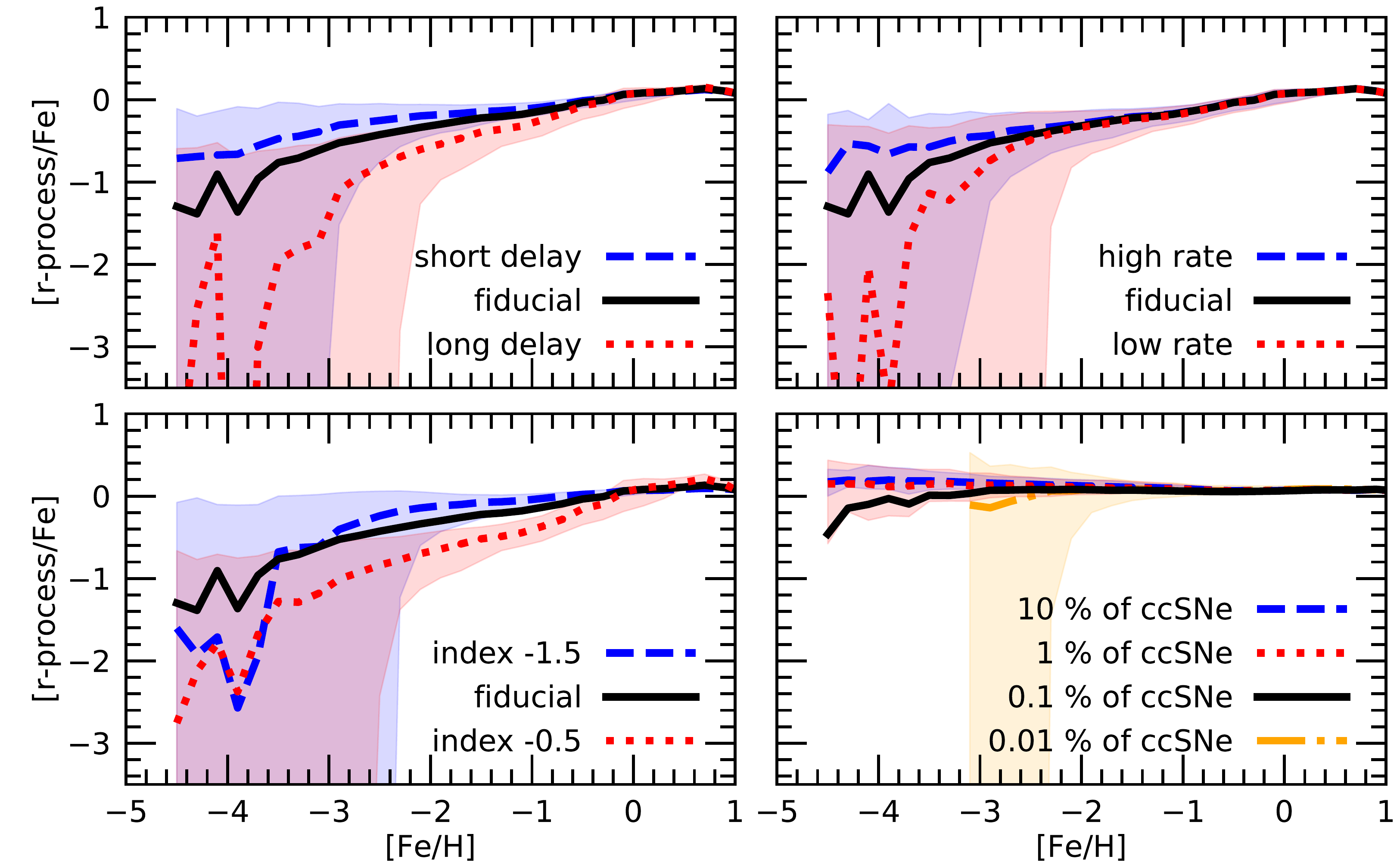}
\caption {\label{fig:var}  $\mathrm{[\rp/Fe]}$ as a function of $\mathrm{[Fe/H]}$ for the different models listed in Tables~\ref{tab:NSmodels} and~\ref{tab:SNmodels}. Curves show the median value and the shaded regions cover the 16th to 84th percentiles of the distribution. From left-to-right and top-to-bottom, the different panels show variations in neutron star merger delay time, neutron star merger rate, power-law time dependence (or steepness) of the neutron star merger delay time distribution, and the rate of rare core-collapse supernovae (a fixed fraction of all core-collapse supernovae). All neutron star models show increasing r-process abundance ratios with increasing metallicity. This relation flattens for models with shorter delay time, higher merger rate, and steeper power-law index and steepens for models with longer delay time, lower merger rate, and shallower power-law index. Enrichment of r-process elements via rare core-collapse supernovae results in almost no dependence of $\mathrm{[\rp/Fe]}$ on $\mathrm{[Fe/H]}$, although our fiducial model (0.1 per cent) shows a small decline of $\mathrm{[\rp/Fe]}$ towards the lowest metallicities. The bottom-right panel shows an additional medium-resolution simulation in orange (dot-dashed curve plus shaded region), which uses a model with only 1 in 10,000 core-collapse supernovae as r-process production sites and exhibits large scatter and a decrease of the median $\mathrm{[\rp/Fe]}$ below $\mathrm{[Fe/H]} = -2$. The scatter is large at low $\mathrm{[Fe/H]}$ for all our neutron star merger models, but only for the rare core-collapse supernova models ($\le0.1$~per cent of core-collapse supernovae).}
\end{figure*}

In order to trace the history of the stars, we check whether the disc and halo stars were formed in the main progenitor of the central galaxy (i.e.\ in-situ) or in an external galaxy (i.e.\ ex-situ). The colour coding in Figure~\ref{fig:NSex} shows the same abundance ratios  as the left-hand panel of Figure~\ref{fig:NS} with the fraction of stars formed ex-situ used as colour coding. The in-situ stars vastly outnumber the ex-situ stars; only 8 per cent of all stars within $R_\mathrm{vir}$ have been accreted. However, 78 per cent of metal-poor stars ($\mathrm{[Fe/H]}<-2$) and 83 per cent of extremely metal-poor stars ($\mathrm{[Fe/H]}<-3$) were formed in other galaxies and later accreted onto the central galaxy or its halo. For stars with $\mathrm{[Fe/H]}<-1$, 61 per cent of stars were accreted, in agreement with observations \citep{Matteo2019}.

As with the formation redshift, there is a strong dependence of ex-situ fraction on metallicity. However, the residual dependence on r-process abundance ratio at fixed metallicity is somewhat less strong. Nevertheless, a larger fraction of r-process enhanced metal-poor stars (as well as of r-process-poor stars) was formed ex-situ compared to all metal-poor stars (e.g.\ 91 per cent for stars with $\mathrm{[Fe/H]}<-2$ and $\mathrm{[\rp/Fe]}>0.7$). This is consistent with observations, which also find that r-process enhanced metal-poor stars were likely accreted \citep{Roederer2018} and with empirical models, which find that a large fraction of r-process enhanced metal-poor halo stars could have originated in now-destroyed dwarf galaxies \citep{Brauer2019}. Our other models (with either neutron star mergers or rare core-collapse supernovae) exhibit similar behaviour, so this result also appears to be robust to variations in our enrichment model. 


Figure~\ref{fig:var} shows the r-process-to-iron abundance ratio for all 10 enrichment models: 7 neutron star merger models and 3 rare core-collapse supernova models. All models are normalized to $\mathrm{[\rp/Mg]}=0$ at $\mathrm{[Mg/H]}=0$. The $1\sigma$ scatter is shown for the models which are presented for the first time in this figure. The top left panel shows neutron star merger models with delay times varying from 10~to 100~Myr. The neutron star merger models vary in rate of mergers from $10^{-6}$~to $10^{-5}$~M$_\odot^{-1}$ in the top right panel. The bottom left panel shows results for steep, fiducial, and shallow neutron star merger delay time distribution, varying from -1.5 to -0.5. 

The different models behave as expected. Shorter delay times, higher merger rates, and steeper delay time distributions result in flatter r-process abundance ratios as a function of metallicity. The reason that, at low metallicity, the median $\mathrm{[\rp/Fe]}$ in the steep delay time model is lower than in the fiducial model is likely because the former produces fewer neutron star mergers overall (see Table~\ref{tab:NSmodels}). The $1\sigma$ scatter remains high in this model, even at the lowest metallicities.

The scatter is larger for the neutron star merger models with long delay times or lower merger rates, because these models have fewer r-process events (and because the r-process abundances in each model are renormalized, these events necessarily have higher r-process yields per event). However, the increase in scatter does not offset the decrease in the median $\mathrm{[\rp/Fe]}$ for metal-poor stars and the 16th percentile of the distribution is higher for models with short delay times or higher merger rates. A model with short minimum delay time, high rate, and steep index would result in the highest $\mathrm{[\rp/Fe]}$ at low metallicity and the flattest trend with metallicity, potentially providing a better match with existing observations (see Section~\ref{sec:obs}).

\subsection{Rare supernovae} \label{sec:SN}

\begin{figure*}
\center
\includegraphics[scale=.45]{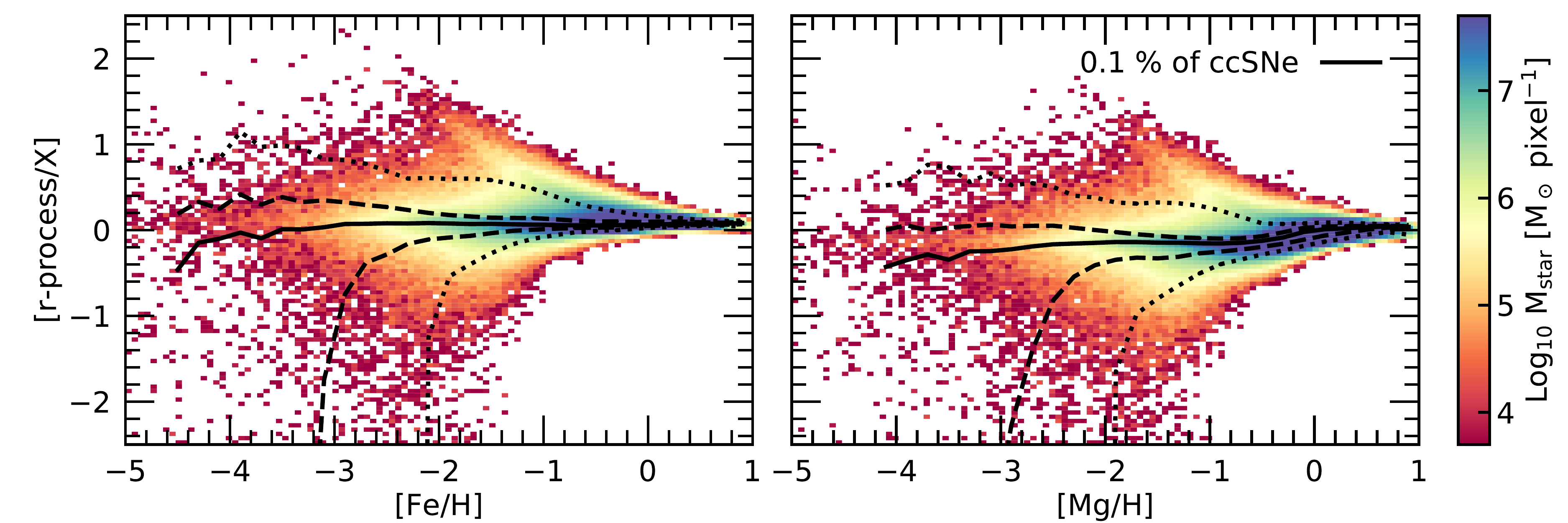}
\caption {\label{fig:SN} The left-hand (right-hand) panel shows $\mathrm{[\rp/Fe]}$ ($\mathrm{[\rp/Mg]}$) as a function of $\mathrm{[Fe/H]}$ ($\mathrm{[Mg/H]}$) for the model in which 1 in 1000 core-collapse supernovae produce r-process elements. The colour coding and curves have the same meaning as in Figure \ref{fig:NS}. The r-process abundance ratios have been normalized so that the median $\mathrm{[\rp/Mg]}=0$ at $\mathrm{[Mg/H]}=0$ (see Table~\ref{tab:SNmodels}). The median r-process abundance ratios are approximately flat, although there is a small increase with increasing metallicity for both iron (at low metallicity) and magnesium (over the full range). The scatter decreases towards high metallicity. There are more extreme outliers than expected from a Gaussian distribution.}
\end{figure*}

Figure~\ref{fig:SN} shows the abundance ratio of r-process elements to iron (magnesium) for the rarest core-collapse supernovae enrichment model (i.e.\ 1 in 1000 core collapse supernovae producing r-process elements) as a function of iron (magnesium) metallicity in the left-hand (right-hand) panel. The median (black, solid curves) varies much less over the full metallicity range, for both $\mathrm{[\rp/Fe]}$ and $\mathrm{[\rp/Mg]}$, than it does for our fiducial neutron star merger model (see~Figure~\ref{fig:NS}). The $1\sigma$ and $2\sigma$ scatter (dashed and dotted black curves) increase towards lower metallicities, but are slightly smaller than for the fiducial neutron star merger model. This is likely due to the fact that r-process element production in rare core-collapse supernova models is more correlated with the production of other metals, because in this case all elements are produced in supernovae. As before, the distribution shows a relatively large number of extreme $\mathrm{[\rp/Fe]}$ outliers. 

The bottom right-hand panel of Figure~\ref{fig:var} shows the abundance ratios for models where the fraction of core-collapse supernovae involved in producing r-process elements decreases from 10 to 1 to 0.1 per cent for the dashed, dotted, and solid curves, respectively. At low metallicity, the median is slightly lower for models with lower rates of r-process events and the scatter increases. For our high-resolution simulation the rarest supernova model has 1 in 1000 core-collapse supernovae producing r-process elements. This model shows a small decrease of $\mathrm{[\rp/Fe]}$ towards very low metallicities and has the largest scatter of the three models (shown in Figure~\ref{fig:SN}).

Additionally, we ran a medium-resolution simulation including a model with only 1 in 10,000 core-collapse supernovae producing r-process elements. This is shown as the orange, dot-dashed curve and shaded region in Figure~\ref{fig:var}. This model results in a lower median $\mathrm{[\rp/Fe]}$ than the other core-collapse supernova models at $\mathrm{[Fe/H]}\lesssim-2$. As expected, the scatter produced by this model is the largest. The median $\mathrm{[\rp/Fe]}$ decreases towards lower metallicities, although not as steeply as the neutron star merger models do. At this extremely low rate of r-process producing events, the r-process site is so rare that it may become difficult to efficiently pollute the ISM at very high redshift, even for prompt core-collapse supernovae, but a high-resolution simulation is necessary to statistically sample below $\mathrm{[Fe/H]}\approx-3$.

The production of iron or magnesium by the same objects that produce r-process elements is higher for the rare core-collapse supernova scenario than for the neutron star merger scenario and therefore more likely to affect the resulting abundance ratios. This would be especially important for metal-poor stars that are enhanced in r-process elements, which could move to higher $\mathrm{[Fe/H]}$ ($\mathrm{[Mg/H]}$) and lower $\mathrm{[\rp/Fe]}$ ($\mathrm{[\rp/Mg]}$), reducing the number of r-process enhanced metal-poor stars. This simultaneous production of iron (magnesium) is not included in our simulations, but should be kept in mind when interpreting our results. However, as shown in Section~\ref{sec:sat}, the r-process elements produced by a single event mix with a substantial amount of pristine and metal-enriched gas. Therefore, the iron production by the same source is likely negligible, at least in our simulations. Detailed yields for different rare core-collapse supernova models, such as magneto-rotational supernovae or collapsars, are necessary to determine whether or not this affects the abundance ratios.

\subsection{Variation between galaxies} \label{sec:diff}

An interesting question is whether different Milky Way-like galaxies have different abundance patterns. We simulated only one galaxy at high resolution, so we are unable to answer this question with our fiducial simulation. However, we ran 16 medium-resolution simulations, all with different initial conditions and therefore producing galaxies with different star formation and merger histories. These simulations do not form enough extremely metal-poor stars to probe this rare population, but their resolution is sufficient for stars with $\mathrm{[Fe/H]} > -3$. We use the same renormalization for all 16 simulations, based on halo L8, which is the galaxy we focus on in this work. This means that we use the same r-process yield per event given a specific enrichment model for each simulation. 

\begin{figure}
\center
\includegraphics[scale=.4]{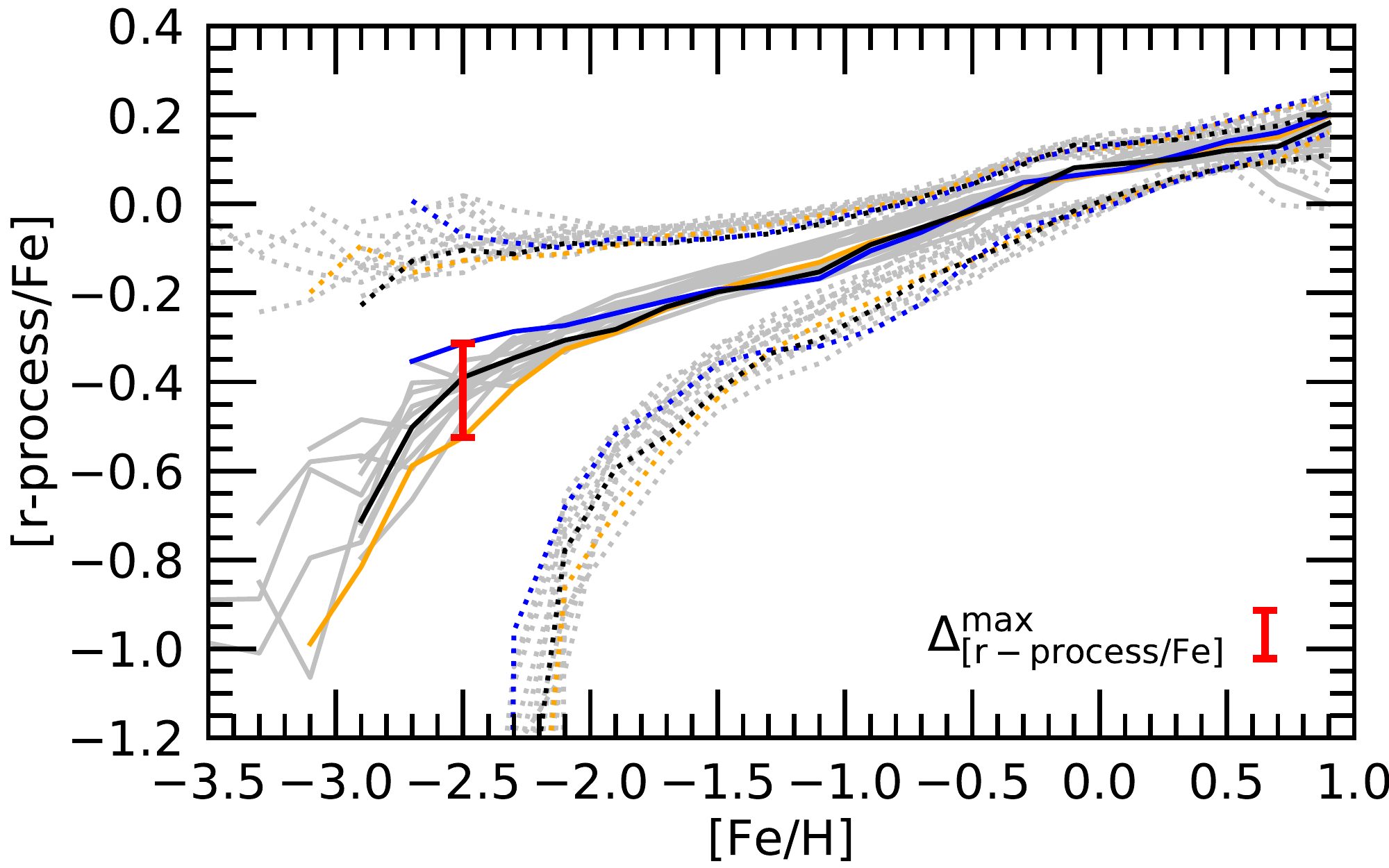}
\caption {\label{fig:diffrpFe} $\mathrm{[\rp/Fe]}$ as a function of metallicity for 16 different Milky Way-mass galaxies and their haloes for the r-process enrichment model with low rate of neutron star mergers. Our fiducial galaxy is shown in black and the two galaxies most discrepant at low metallicity are shown as blue and orange curves. The median ($1\sigma$ scatter) is shown by the solid (dotted) curves. The abundance ratios are reasonably similar for all our galaxies, each showing an increase in $\mathrm{[\rp/Fe]}$ with increasing metallicity. Our fiducial galaxy has relatively average abundance ratios. The variation between galaxies increases towards low metallicity, where the r-process production sites become extremely rare. The maximum difference at $\mathrm{[Fe/H]}=-2.5$ is shown by the red error bar and listed for all r-process models in Table~\ref{tab:diff}.}
\end{figure}

In Figure~\ref{fig:diffrpFe} we show the abundance ratios for all 16 galaxies for one of the r-process enrichment models with relatively large variation between galaxies (model ``low rate''). The abundances for the galaxy with the highest (lowest) $\mathrm{[\rp/Fe]}$ at $\mathrm{[Fe/H]}=-2.5$ is shown in blue (orange) and fiducial halo L8 is shown in black. Solid curves show the median values and dotted curves the 16th and 84th percentiles. 

At the metallicities we can reliably probe with our medium-resolution simulations, the maximum difference in $\mathrm{[\rp/Fe]}$ between our 16 simulations is less than 0.3 dex. Additionally, our fiducial halo L8 is an average halo in terms of its r-process abundances (see Figure~\ref{fig:diffrpFe}). We therefore conclude that all our results based on a single high-resolution simulation are likely to hold for any Milky Way-like galaxy. However, the variation between galaxies tends to increase towards lower metallicity \. There could be larger differences in the extremely metal-poor regime, because fewer events contributed to the r-process enrichment, but a suite of high-resolution simulations is needed to test this. When the source(s) of r-process elements are known and understood, these rare elements could potentially help constrain the early star formation and enrichment history of the Milky Way. 

We find that exactly how similar r-process abundance ratios are in different galaxies depends on the specific model for r-process enrichment. For example, our fiducial neutron star merger model shows negligible differences between galaxies. However, for the model with low rate of neutron star mergers, the abundance ratios diverge somewhat towards the low-metallicity end, as shown in Figure~\ref{fig:diffrpFe}.

\begin{table}
\begin{center}                                                                                                                                        
\caption{\label{tab:diff} \small Variation of r-process abundance in low-metallicity stars between different galaxies: model name, model type, the difference between minimum and maximum value of median $\mathrm{[\rp/Fe]}$ at $\mathrm{[Fe/H]} = -2.5$ for 16 different Milky Way-mass galaxies (shown in Figure~\ref{fig:diffrpFe} for model ``low rate''), and the difference at $\mathrm{[Fe/H]} = -1.5$.}         
\begin{tabular}[t]{llcc}
\hline \\[-3mm]                                                                                                                                       
  model name   & model type  & $\Delta^\mathrm{max}_\mathrm{[\rp/Fe]}$& $\Delta^\mathrm{max}_\mathrm{[\rp/Fe]}$ \\
               &   & $\mathrm{[Fe/H]} = -2.5$ & $\mathrm{[Fe/H]} = -1.5$ \\
\hline \\[-4mm]                                                                                                                                       
  fiducial NS     &  NS merger & 0.105 & 0.078 \\
  high rate       &  NS merger & 0.140 & 0.078 \\
  low rate         &  NS merger & 0.211 & 0.071 \\
  short delay    &  NS merger & 0.092 & 0.047 \\
  long delay     &  NS merger & 0.257 & 0.170 \\
  shallow DTD  &  NS merger & 0.225 & 0.151 \\
  steep DTD     &  NS merger & 0.229 & 0.055 \\
  high fraction       & rare ccSNe   & 0.030 & 0.036 \\
  medium fraction & rare ccSNe   & 0.040 & 0.034 \\
  low fraction        & rare ccSNe & 0.045 & 0.030 \\
  extra low fraction & rare ccSNe & 0.252 & 0.045 \\
\hline                                                                                                                                                
\end{tabular}                                                                                                                                         
\end{center}                                                                                                                                          
\end{table}      

We quantify this for all r-process models in Table~\ref{tab:diff} by calculating the difference between the simulation with the highest median $\mathrm{[\rp/Fe]}$ and one with the lowest median $\mathrm{[\rp/Fe]}$ at fixed metallicity of $\mathrm{[Fe/H]} = -2.5$ (as indicated by the red error bar in Figure~\ref{fig:diffrpFe}) and at $\mathrm{[Fe/H]} = -1.5$. We use our 10 standard models plus the additional model (``extra low fraction'') with 1 in 10,000 supernovae producing r-process elements. For models where the source of r-process elements is more rare (and the yield per event is higher), the scatter between different galaxies at low metallicity is larger. At higher metallicities, $\mathrm{[Fe/H]} \ge -2$, the maximum difference between simulations is below 0.1~dex for all models except the one with longer minimum delay time and the one with a shallower delay time distribution, because these two models are less correlated with the star formation rate of the galaxy.

\subsection{Convergence} \label{sec:res}

\begin{figure}
\center
\includegraphics[scale=.4]{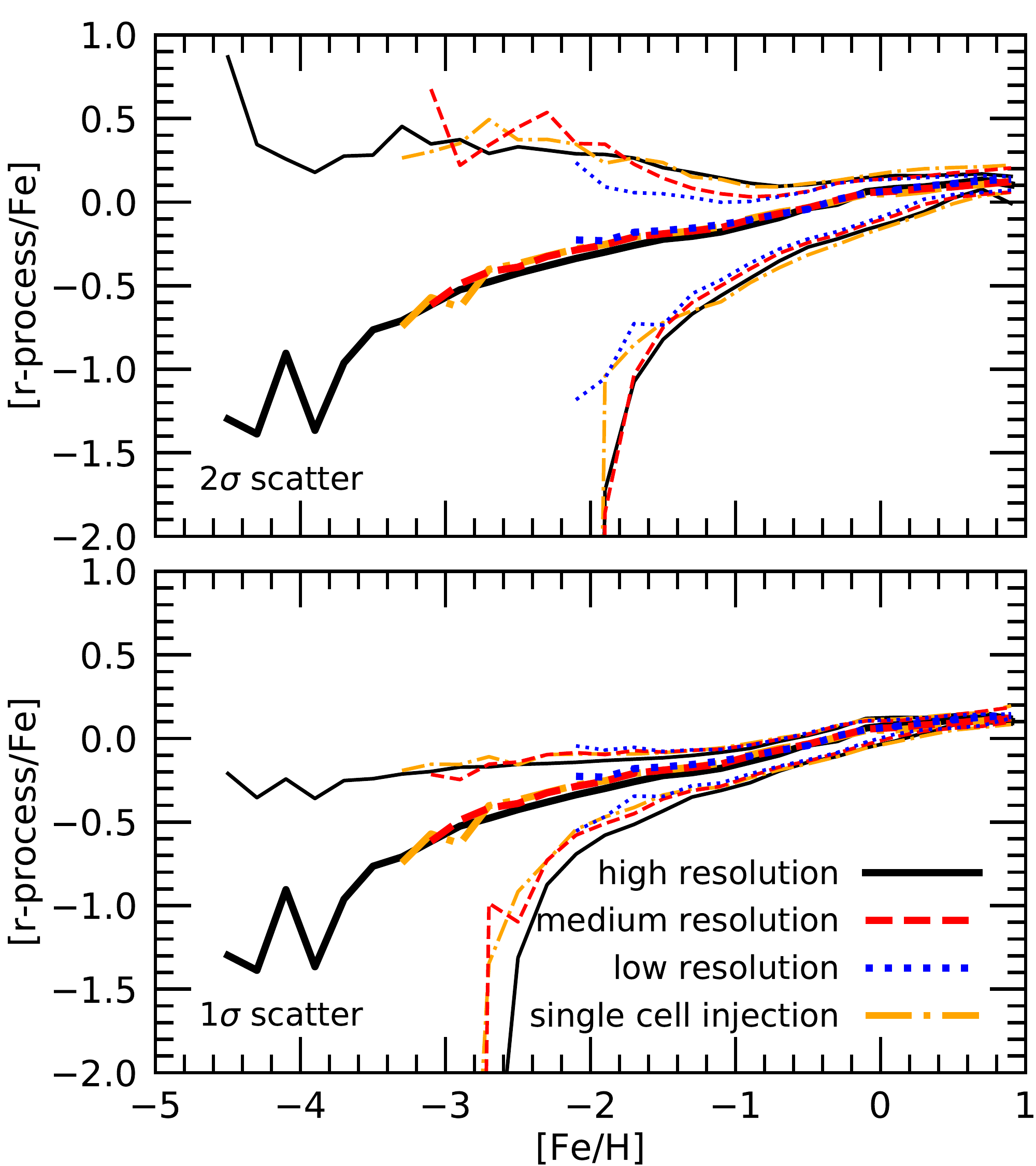}
\caption {\label{fig:res} Median (thick curves) and 98th and 2nd percentiles (top panel, thin curves) and 84th and 16th percentiles (bottom panel, thin curves) of $\mathrm{[\rp/Fe]}$ as a function of $\mathrm{[Fe/H]}$ for our fiducial neutron star merger enrichment model using the normalization of the high-resolution simulation (black solid curves). Simulations with 8 and 64 times lower resolution are shown as red dashed (medium resolution) and blue dotted curves (low resolution), respectively. A simulation with medium resolution, but an initial enrichment region consisting of a single gas cell (the host cell) instead of 64 neighbouring cells is shown by the orange, dot-dashed curves. Only bins containing at least 100 star particles are shown. The simulations are well-converged and the choice of initial enrichment region does not affect our conclusions. The scatter increases slightly with increasing resolution. This increase is larger for the $2\sigma$ scatter than for the $1\sigma$ scatter. Reducing the injection region of mass and metals ejected by star particles only slightly enhances the $2\sigma$ scatter, but does not noticeably affect the $1\sigma$ scatter of this model.}
\end{figure}

Previous particle-based cosmological simulations that followed r-process enrichment, but did not allow for the exchange of mass and metals between resolution elements, found strong dependence on resolution \citep{Voort2015a}. At $\mathrm{[Fe/H]}=-3$ the r-process abundance dropped by more than an order of magnitude when increasing the mass resolution by a factor of~8. This was due to the fact that the initial enrichment region around a neutron star merger decreased with increasing resolution in combination with the complete absence of mass and metal exchange between gas particles. The newly produced r-process elements were distributed over a fixed number of neighbouring gas particles, rather than over a fixed mass or volume. Metals obtained were stuck to their initial resolution element and not able to mix with neighbouring gas particles. 

In a very similar way, the moving mesh simulations in this work also distribute heavy elements over a fixed number of neighbouring gas cells, which means that the initial enrichment region decreases with increasing resolution as well. The grid cells in our calculations move with the bulk of the flow as in particle-based simulations. However, the gas, including the metals, can subsequently move to and mix with different resolution elements, which means that the choice of initial enrichment region is much less important. Note that the mixing between cells is treated self-consistently, as determined by the local velocity field, and not imposed by any subgrid mixing model. This small-scale mixing removes the dependence on resolution and on parameter values chosen for the initial enrichment, as shown in Figure~\ref{fig:res}. Black, solid curves show the median and $2\sigma$ (top panel) and $1\sigma$ (bottom panel) scatter in $\mathrm{[\rp/Fe]}$ in our high-resolution simulation for our fiducial neutron star merger model. Additionally, the red, dashed and the blue, dotted curves show the same quantities for simulations with medium and low resolution, respectively. We use the normalization of the high-resolution simulation for all simulations shown, which means the same r-process yield per event is used for each simulation. 
  
All simulations show very similar r-process abundances, with up to 0.1 dex difference in the median value between simulations that vary a factor 64 in mass resolution. We therefore conclude that the metal enrichment in our simulation is well-converged. The scatter increases slightly at higher resolution, especially the $2\sigma$ scatter. Although this is a small effect, the initial enrichment region is smaller and numerical mixing is reduced at higher resolution, which could both contribute to producing slightly more extreme outliers. The scatter increases somewhat more strongly with resolution for our fiducial rare core-collapse supernova model than for the fiducial neutron star merger model shown here, likely because of the stronger correlation between star formation and r-process enrichment. When considering only 10 star particles per metallicity bin instead of the 100 we use throughout this work, the result is noisy in the lowest-metallicity regime and the convergence behaviour is worse. This is the reason that we only trust and show results for bins with at least 100 star particles. 

\begin{figure*}
\center
\includegraphics[scale=.49]{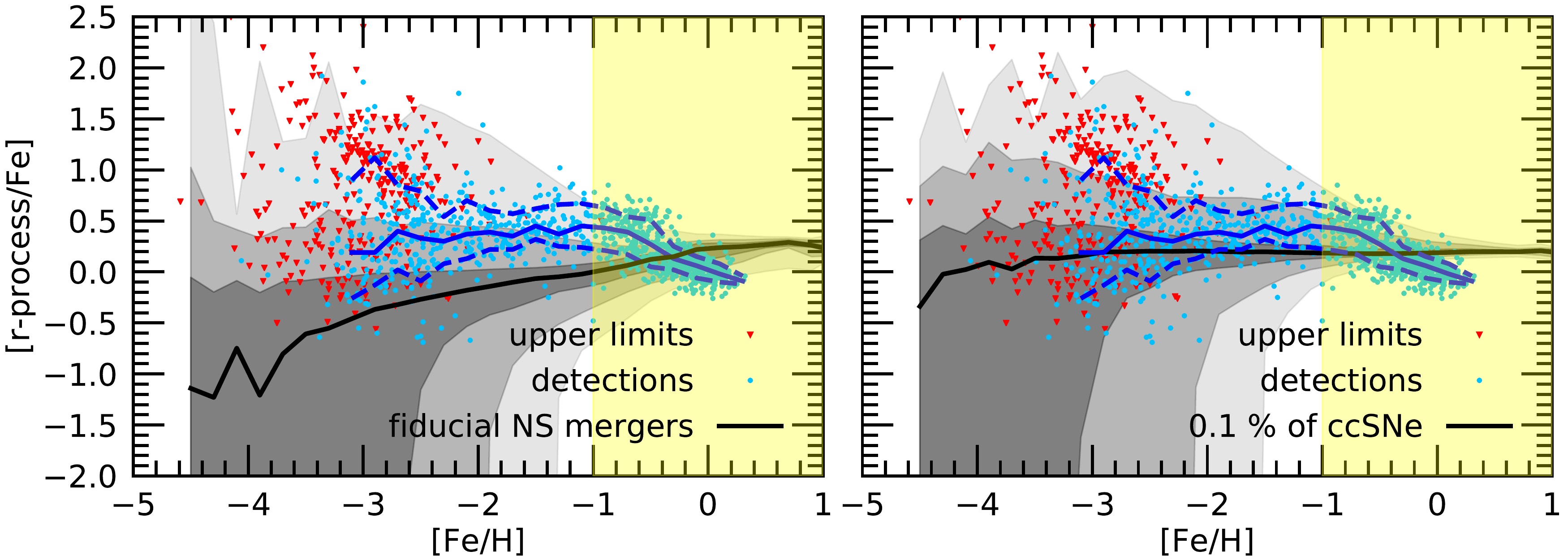}
\caption {\label{fig:obs} Median (black curves) and $1\sigma$, $2\sigma$, and $3\sigma$ scatter (grey shaded regions) of $\mathrm{[\rp/Fe]}$ as a function of $\mathrm{[Fe/H]}$, where the r-process abundances are normalized to $\mathrm{[\rp/Mg]}=0$ at $\mathrm{[Mg/H]}=0$ (see Tables~\ref{tab:NSmodels} and~\ref{tab:SNmodels} for the yields used). The left-hand panel shows results from our fiducial neutron star merger model, while the right-hand panel shows r-process enrichment from 1 in 1000 core-collapse supernovae. Measurements of [Eu/Fe] in Milky Way stars are shown as light blue circles (detections) and red downward triangles (upper limits) from the database compiled by \citet{Suda2008}. The blue solid and dashed curves show the median and $1\sigma$ scatter of the observational detections. We exclude the yellow shaded region ($\mathrm{[Fe/H]}>-1$) from our comparison. The different trends with metallicity for the two models seem to suggest that observations prefer rare core-collapse supernovae over neutron star mergers as the dominant source of r-process elements.}
\end{figure*}

To understand the importance of our choice to initially mix the newly produced metals into 64 neighbouring cells when they are ejected by a star particle, we ran an additional medium-resolution simulation in which we change the initial injection region to only the single host gas cell of the star particle. This is shown by the orange, dot-dashed curves. For our fiducial neutron star merger model, the median and $1\sigma$ scatter are unaffected by this factor of 64 change in the initial gas mass enriched by an r-process event. Our fiducial core-collapse supernova model exhibits a slight increase in the $1\sigma$ scatter, because r-process enrichment traces star formation more closely, but the median is also unaffected. This allows us to conclude that our results are robust to changes in the injection region. The $2\sigma$ scatter does increase slightly, showing that it is the extreme outliers (both high and low) which are most affected by the initial enrichment.

The metallicity distribution function is converged at subsolar metallicities and also does not depend on the choice of initial enrichment region. For every dex decrease in $\mathrm{[Fe/H]}$, there are about an order of magnitude fewer stars. This is shown in Appendix~\ref{sec:FeMg}.

The convergence problem for metal abundances in fully Lagrangian codes can likely be remedied by adding an appropriate subgrid metal diffusion model to the hydrodynamics \citep[e.g.][]{Shen2010, Pilkington2012, Escala2018}. These models have been shown to reduce the number of extremely metal-poor stars and the scatter in abundance ratios at fixed metallicity. Currently, such efforts have not yet included on-the-fly r-process enrichment. In the future, it would be very interesting to compare galactic r-process abundances obtained with our moving mesh technique to Lagrangian methods that include subgrid metal diffusion.

\section{Comparison with observations} \label{sec:obs}

Europium is almost completely produced by the r-process \citep{Burris2000} and has been observed in the Milky Way and in surrounding dwarf galaxies across a wide range of metallicities. It is therefore an ideal observational tracer of r-process enrichment. In this section, we will compare our simulated r-process abundance ratios to observed $\mathrm{[Eu/Fe]}$, but still refer to it as $\mathrm{[\rp/Fe]}$. Observational data of stellar abundances in the Milky Way, its halo, and its satellites were taken from the SAGA database, compiled by and described in \citet{Suda2008}. Solar abundances for both observations and simulations are taken from \citet{Asplund2009}. We use only metal-poor and extremely metal-poor stars from the SAGA database and exclude carbon-enhanced stars, as also done by \citet[e.g.][]{Wehmeyer2019}. This results in a total sample of 1388 stars. We checked that if we restrict the sample to only include stars observed with a spectral resolution of $R>10,000$, resulting in 1217 stars, the distribution is very similar and none of our conclusions are changed. Observational selection effects and errors are not modelled in our simulated data. It is therefore not possible to compare our models and available observations in detail, but still enlightening to look at general trends. 

As discussed before, the insufficient iron production by Type Ia supernovae results in unrealistic abundance ratios at high metallicity ($\mathrm{[Fe/H]}>-1$) and we therefore cannot compare our simulations to observations in this regime. Instead, we focus on the low-metallicity stars. As in the previous section, we normalize the r-process yields to result in a median $\mathrm{[\rp/Mg]}$ of zero at $\mathrm{[Mg/H]}=0$, because our simulated abundances do not show the $\approx0.4$~dex decrease in $\mathrm{[\rp/Fe]}$ (and $\mathrm{[\alpha/Fe]}$) towards high metallicity. The resulting yields are listed in Tables~\ref{tab:NSmodels} and~\ref{tab:SNmodels}.

\subsection{Milky Way} \label{sec:MW}


Abundances from our fiducial neutron star merger enrichment model are shown in the left-hand panel of Figure~\ref{fig:obs} and those from our fiducial core-collapse supernova enrichment model are shown in the right-hand panel. The grey shaded regions show the $1\sigma$, $2\sigma$, and $3\sigma$ scatter in bins with at least 100 star particles. Different symbols show the observational data, where light blue circles show europium detections and red downward triangles show upper limits. The blue solid and dashed curves show the median and $1\sigma$ scatter of the observational detections.

The largest difference between r-process enrichment models through neutron star mergers and those through rare core-collapse supernovae is the dependence of the median $\mathrm{[\rp/Fe]}$ on metallicity. Neutron star merger models show a modest, but significant, increase, whereas the median $\mathrm{[\rp/Fe]}$ in rare core-collapse supernova models is almost independent of metallicity. It is important to note that the observational data are relatively sparse for extremely metal-poor stars and potentially biased due to unknown selection effects in the heterogeneous sample, which are not taken into account. 

Although the normalization of our simulations is somewhat uncertain, the trend with metallicity appears to be steeper in our fiducial neutron star merger model than in the observational data. A higher rate of neutron star mergers, shorter delay time, and steeper delay time distribution would improve the match with observations of low-metallicity stars. In contrast to the neutron star merger models, the rare core-collapse supernova model matches the observational data reasonably well at $\mathrm{[Fe/H]}<-1$, although the scatter is slightly too small. A model with a lower event rate (as shown in Figure~\ref{fig:var}) could improve the agreement, but this needs to be tested with a new high-resolution simulation. When comparing the median trend with metallicity, currently available observations seem to prefer rare core-collapse supernovae over neutron star mergers as the dominant source of r-process element, although both source types may contribute. We caution the reader, however, that a bigger and unbiased observational sample is necessary to rule out any of our r-process enrichment models. We discuss possible future improvements to our simulations in more detail in Section~\ref{sec:concl}.

At $-2<\mathrm{[Fe/H]}<-1$, the observations exhibit larger scatter than our simulations. At $\mathrm{[Fe/H]}<-2$, the scatter in $\mathrm{[\rp/Fe]}$ increases towards low metallicity for both models and observations, but the simulations appear to still underproduce the scatter in this regime. This is likely at least in part due to additional observational errors, which are not included in our simulation results. Another possibility is that the r-process production source in reality is more rare than in our fiducial models, which would also increase the scatter (see Figure~\ref{fig:var}). This would, however, also decrease the median $\mathrm{[\rp/Fe]}$ and result in more r-process-free metal-poor stars. Alternatively, a smaller simulated scatter may indicate that the mixing of metals in the ISM is too efficient in our simulations. Even though our simulation results are almost independent of resolution, the scatter increases slightly with improved resolution, a trend which could continue towards even better resolution.

\subsection{Dwarf galaxies} \label{sec:sat}

Interesting differences exist between r-process abundances in the Milky Way’s halo and in its satellites (e.g.\ \citealt{Tolstoy2009, Frebel2018}, and references therein). Because of the rarity of the production site(s) of r-process elements, extremely low-mass galaxies only experience a single or very few r-process-producing events over their lifetime. The stochasticity of these events therefore impacts dwarf galaxies even more strongly than the Milky Way. A sufficiently low-mass dwarf galaxy will experience either no r-process enrichment if no events occur or strong r-process enhancement if one event occurs. This is the interpretation for the widely different r-process abundances found in local ultra-faint dwarf galaxies, where most have no detected europium, but some of them are strongly r-process enhanced \citep{Ji2016a, Hansen2017}. At high redshift, hydrodynamical simulations have shown that a single neutron star merger could indeed lead to high r-process abundances in very low-mass galaxies \citep{Safarzadeh2017}. 

The limited resolution of our simulation prohibits a study of ultra-faint dwarfs, which have stellar masses $\lesssim10^5$~M$_{\astrosun}$. However, it does allow us to study reasonably low-mass dwarfs with stellar masses around $10^6$~M$_{\astrosun}$ orbiting the central galaxy as satellites. We focus on the r-process enrichment model which results in fewer events over the full history of the satellite galaxies (model ``low rate''). A model with fewer r-process producing events will result in more scatter between different satellites. The neutron star merger rate in our fiducial model, on the other hand, is high enough for satellites of $10^6$~M$_{\astrosun}$ to follow the average enrichment trend (similar to the central galaxy). The variation in r-process abundance ratios for different observed satellites can thus help constrain how rare the site for r-process production is \citep{Beniamini2016}.

\begin{figure}
\center
\includegraphics[scale=.4]{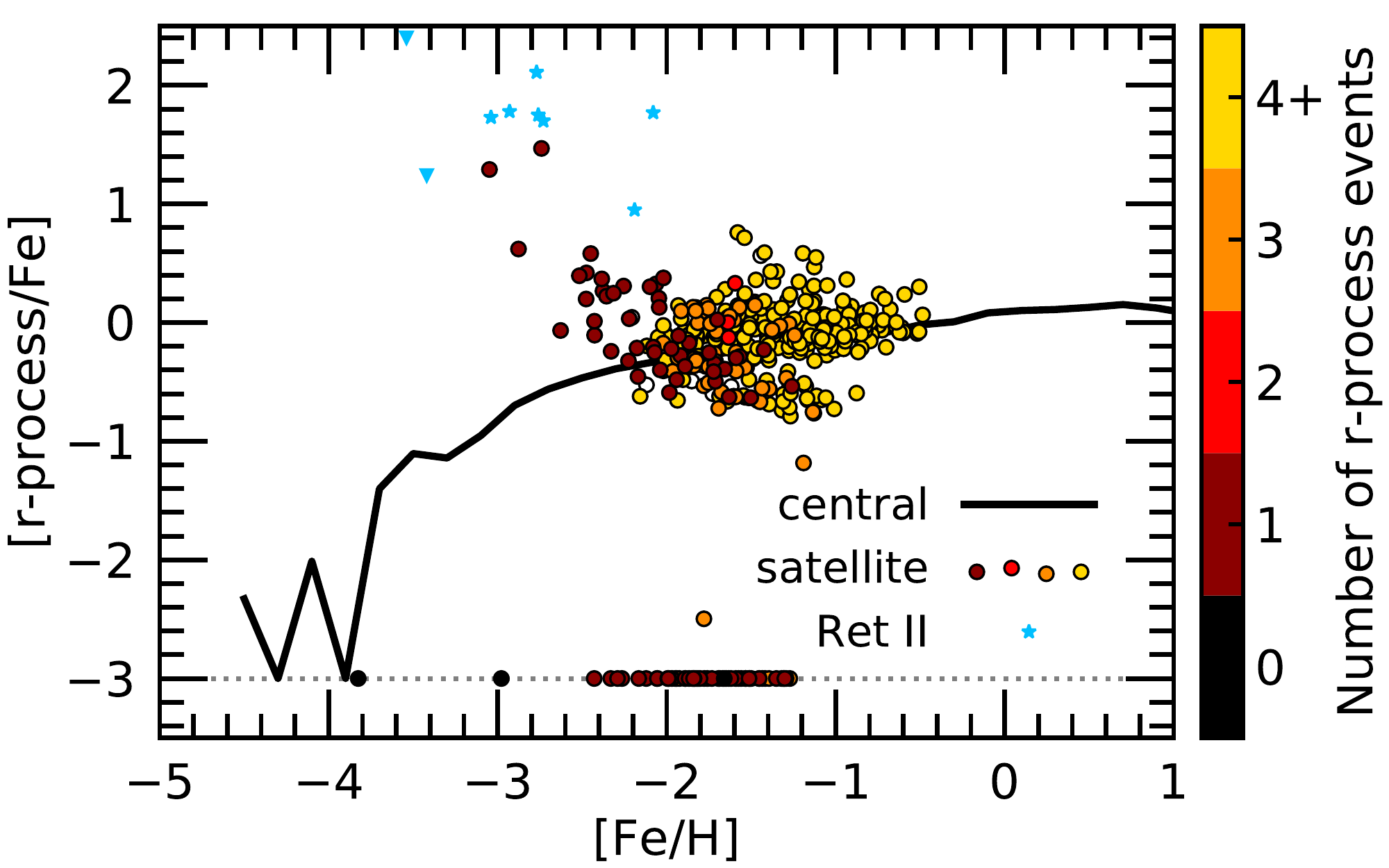} 
\caption {\label{fig:sat} The stellar abundances of individual star particles in a simulated satellite with stellar mass of $10^{6.4}$~M$_{\astrosun}$ are shown as circles and colour coded by how many neutron star mergers the satellite had experienced at the time the stars were born. Star particles for which the number of neutron star mergers was not known exactly are coloured white. Star particles without any r-process enhancement have been set to $\mathrm{[\rp/Fe]}=-3$, as shown by the dotted grey line. The r-process enrichment model used here is the one with a low rate of neutron star mergers (model ``low rate'' in Table~\ref{tab:NSmodels}) and the simulated r-process abundances have been normalized to $\mathrm{[\rp/Mg]}=0$ at $\mathrm{[Mg/H]}=0$, as before. The black curve shows the median abundance ratios for the central Milky Way-mass galaxy. Star symbols (triangles) show detections (upper limits) in 9 (2) metal-poor stars observed in ultra-faint dwarf galaxy Reticulum II \citep[Ret II;][]{Ji2016a}. Our simulated dwarf galaxy experiences an early r-process enrichment event and the stars that form shortly afterwards are strongly enhanced in r-process elements, as also seen in the lower mass observed satellite.} 
\end{figure}
In Figure~\ref{fig:sat}, we show abundance ratios of individual star particles for one of the satellites in our simulation with present-day stellar mass $10^{6.4}$~M$_{\astrosun}$. The black, solid curve shows the median abundance ratio for the central galaxy, as shown before in Figure~\ref{fig:var}. The star particles are colour coded by the number of neutron star mergers that occurred in the dwarf galaxy before the star was born. To do this we need to include the stars that were stripped from the dwarf galaxy as well. We use merger trees \citep{Springel2005} to trace the progenitor galaxy and identify all the stars that used to belong to the galaxy. We then use all stars (including stripped stars) to calculate the cumulative number of r-process events as a function of time. Each star particle keeps track of how often it experienced an r-process event, which is saved at 128 output times. This limits our time resolution, so for a small fraction of stars we do not know whether the r-process event took place before or after the star formed. In this case, the circles in Figure~\ref{fig:sat} are coloured white. For visualization purposes, r-process-free stars have been set to $\mathrm{[\rp/Fe]}=-3$ (as also indicated by the grey, dotted line). Star symbols and triangles show detections and upper limits of metal-poor stars observed in ultra-faint dwarf galaxy Reticulum II \citep{Ji2016a}. The observations and simulations are not directly comparable, because of the different mass regime.

The satellite shown in Figure~\ref{fig:sat} experienced a very early r-process enrichment event (at $z>9.4$). This results in a large fraction of the metal-poor stars being strongly r-process enhanced, especially at $\mathrm{[Fe/H]}\approx-3$. This dwarf was specifically selected to show this behaviour and there are also examples of dwarfs in our simulations which show no r-process enrichment in metal-poor stars. Such differences emphasize the strong stochastic nature of r-process enrichment in our model. This is qualitatively similar to the observed variation between ultra-faint dwarfs, although at much higher stellar mass. If the r-process enrichment site is indeed as rare as in our rarest models (i.e.\ a $z=0$ rate of less than $10^{-4}$~yr$^{-1}$), as also suggested based on ultra-faint dwarf abundances \citep{Beniamini2016}, then our simulations predict that (extremely) metal-poor stars in classical dwarfs will also show large r-process abundance variations, similar to those in ultra-faint dwarfs.  

Besides strongly r-process enhanced stars, we furthermore find a substantial r-process-free population within the satellite shown in Figure~\ref{fig:sat}. Stars without any r-process elements were still forming after the first r-process enrichment event. We traced this satellite back in time and found that it underwent a major merger at $z\approx2$ in which one of the partners had experienced an early neutron star merger (at $z>9.4$) and the other partner had its first neutron star merger at a much later time $z<3.5$. When these two dwarf galaxies merged (before becoming a satellite), the r-process enhanced population of one galaxy and the mostly r-process-free population of the other galaxy were combined, which results in the bimodal distribution seen at $z=0$. 

\begin{figure}
\center
\includegraphics[scale=.42]{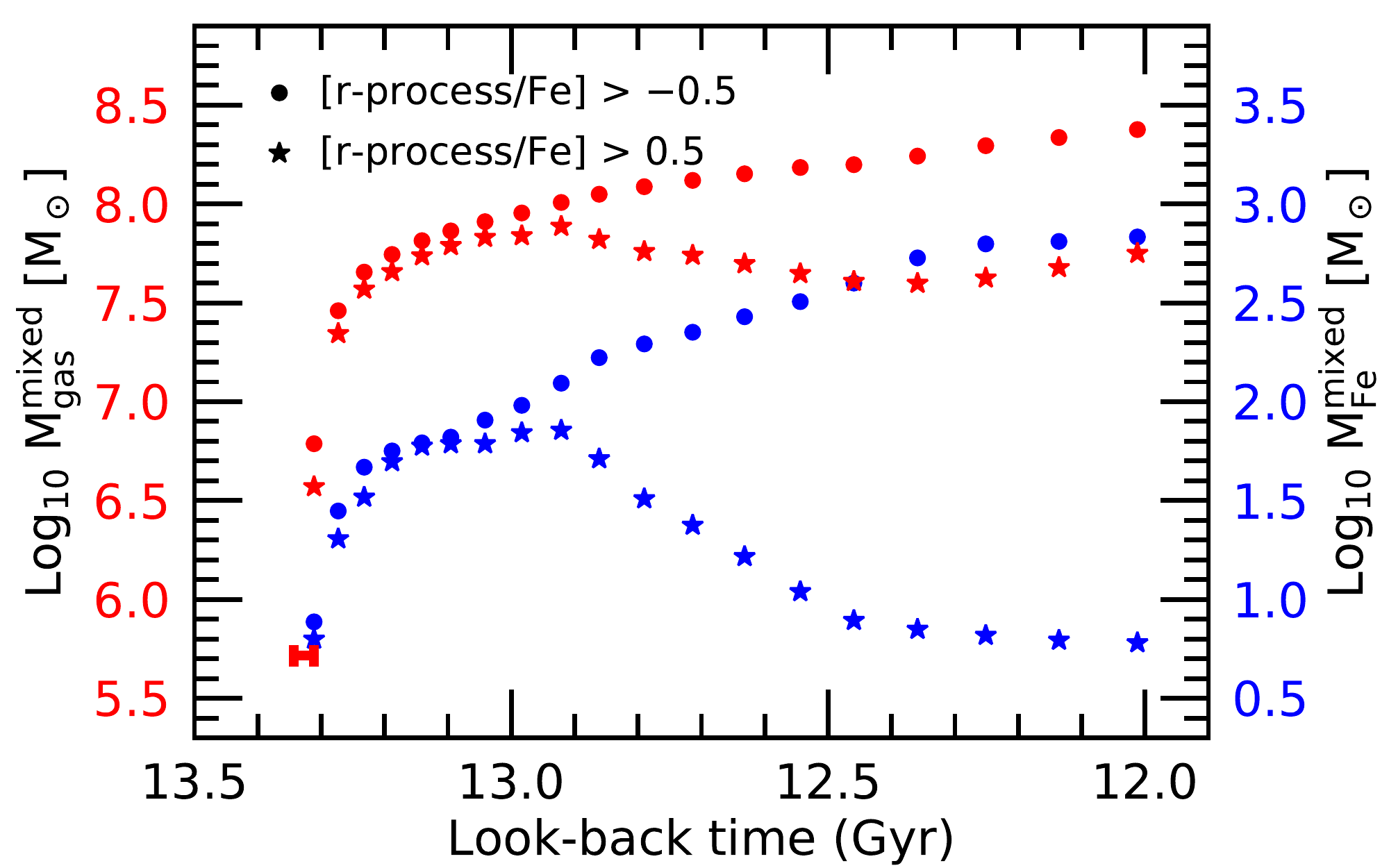} 
\caption {\label{fig:mix} The total amount of gas (left $y$-axis) enriched to $\mathrm{[\rp/Fe]}>-0.5$ (red circles) and $\mathrm{[\rp/Fe]}>0.5$ (red stars) by a single r-process event in the dwarf galaxy shown in Figure~\ref{fig:sat}, using the ``low rate'' neutron star merger enrichment model, as a function of look-back time. The total amount of iron (right $y$-axis) mixed with the r-process enriched gas is shown by the blue symbols. Initially, the r-process event enriches 64 gas cells with newly produced heavy elements, which is shown by the red error bar. The r-process material quickly mixes with about $10^8$~M$_{\astrosun}$ of gas and $10^2$~M$_{\astrosun}$ of iron in the first few hundred~Myr and continues to mix with more gas over the next~Gyr.} 
\end{figure}

The stellar abundance ratios depend, in part, on the amount of mixing of the gas from which they formed. If the r-process ejecta mix with mostly pristine gas, then $\mathrm{[\rp/Fe]}$ stays approximately constant, while $\mathrm{[Fe/H]}$ decreases. If the r-process enhanced gas mixes with metal-rich (i.e.\ iron-rich) gas, then $\mathrm{[\rp/Fe]}$ decreases and $\mathrm{[Fe/H]}$ increases. 

In Figure~\ref{fig:mix} we show the evolution of the total gas mass (left $y$-axis) enriched by a single neutron star merger to values of $\mathrm{[\rp/Fe]}>-0.5$ (red circles) and $\mathrm{[\rp/Fe]}>0.5$ (red stars), for the same dwarf galaxy and r-process enrichment model as shown in Figure~\ref{fig:sat}. Shown in blue, on the right $y$-axis, is the amount of iron mixed with the r-process enriched gas. Initially, the r-process event enriches 64 gas cells with newly produced heavy elements, which equates to $M_\mathrm{gas}^\mathrm{enriched}=10^{5.7}$~M$_\odot$ for our high resolution simulation and is shown by the red error bar. The r-process event occurs after $z=9.8$, but before $z=9.4$, which is the time between two consecutive simulation outputs. No other neutron star merger occurs within a 40~kpc radius for the next 1.3~Gyr (until $z<3.5$). We can therefore use this event to investigate the amount of mass that becomes enriched with r-process elements in and around this dwarf galaxy due to a singular r-process event.

In our simulation, the r-process material quickly mixes with about $10^8$~M$_{\astrosun}$ of gas and $10^2$~M$_{\astrosun}$ of iron in the first few hundred~Myr. After this initial phase, the amount of gas enriched to $\mathrm{[\rp/Fe]}>0.5$ decreases, due to the continued production of iron in the dwarf galaxy, which mixes with the r-process enriched gas. On the other hand, the amount of gas enriched to $\mathrm{[\rp/Fe]}>-0.5$ continues to increase, because the r-process enriched region continues to expand to several tens of kpc, far beyond the dwarf galaxy's radius, and thus continues to mix with more pristine gas. Most of the gas has $\mathrm{[\rp/Fe]}>-0.5$ for this part of the evolution, so if we lowered the threshold to e.g.\ $\mathrm{[\rp/Fe]}>-2$, the evolution would be practically identical. 

This is just one example of the (iron) mass mixed in with that of a single r-process event. The amount of mixing could be somewhat different in more massive galaxies and in the present-day Universe. Additionally, the ISM model used does not capture its true multi-phase nature, as also discussed below. However, this example shows that the amount of mixing in our simulations is reasonably high. Our values are consistent with those derived by \citet{Macias2019}, who find high levels of mixing could be necessary to match observed abundances of extremely metal-poor stars. It may also imply that the production of iron in r-process producing core-collapse supernovae, which is not included in our simulations, is negligible compared to the iron mass produced by other supernovae and mixed in with the r-process enriched gas.

\section{Comparison with previous work} \label{sec:prev}

This work is the first time r-process elements are produced and traced on-the-fly in high-resolution, cosmological, moving mesh simulations. It is useful to compare our newly obtained results for r-process enrichment by neutron star mergers to previous work, which used different hydrodynamical techniques and/or post-processed their simulation. All cosmological simulations that include models for r-process enrichment by neutron star mergers find that the majority of metal-poor stars are enriched with r-process elements \citep{Shen2015, Voort2015a, Naiman2018}, unlike more idealized chemical evolution models \citep[e.g.][]{Argast2004, Matteucci2014, Wehmeyer2015}, and that the scatter in $\mathrm{[\rp/Fe]}$ is large. This result therefore appears to be robust and shows that neutron star mergers are likely relevant for enriching (extremely) metal-poor stars with r-process elements, although it remains unclear whether or not they dominate. However, the details of the r-process abundances can be quite different between the various cosmological simulations.

The highest resolution particle-based simulation in \citet{Voort2015a} also showed an increasing trend of $\mathrm{[\rp/Fe]}$ with metallicity, as we do in this work. This simulation is therefore in qualitative agreement with ours. However, their r-process abundances were not converged due to the lack of mixing at the resolution scale. Their fiducial (medium resolution) simulation resulted in an approximately flat $\mathrm{[\rp/Fe]}$ trend with metallicity, and a low resolution simulation even showed a decreasing trend. Therefore, there is some uncertainty as to whether the metallicity trends in \citet{Voort2015a} are physically or numerically driven and thus to what extent our results agree. The addition of a subgrid turbulent diffusion model would likely help to improve the convergence properties and reveal a more robust trend with metallicity \citep[e.g.][]{Shen2010, Pilkington2012, Escala2018}. This work also includes a range of neutron star merger model parameters, finding qualitatively similar differences between models as we do in our simulations. However, the variation between models in \citet{Voort2015a} is larger than the variation between our models, which can likely also be attributed to the lack of small-scale mixing.

\citet{Shen2015} used a post-processing enrichment approach, which means they injected r-process elements after the simulation was finished. The results could therefore depend on the spacing between output times of the simulation. Their fiducial model additionally includes a metallicity floor for iron and a subgrid mixing scheme (assigning abundances based on 128 neighbouring gas particles) for both iron and r-process elements. This study found that the median $\mathrm{[\rp/Fe]}$ decreases with increasing metallicity, as opposed to the results presented in this work. Unfortunately, whether or not these results depend on resolution is not discussed. We believe the discrepancy with our simulations is likely due to the post-processing approach in \citet{Shen2015} and it would be interesting to see how our results compare when both simulations apply on-the-fly r-process enrichment.

\citet{Naiman2018} used moving mesh simulations with on-the-fly r-process enrichment, which included a metallicity floor for both iron and r-process elements, and a subgrid scheme in post-processing to redistribute r-process elements to only part of the mass contained by a star particle. The latter was necessary because of the limited resolution of the simulation (owing to the much larger volume simulated, yielding many Milky Way-mass galaxies). This study also found that the median $\mathrm{[\rp/Fe]}$ decreases with increasing metallicity, again contrary to our current results. We believe the imposed metallicity floor in \citet{Naiman2018} is responsible for this behaviour, because it artificially increases $\mathrm{[\rp/Fe]}$ at $\mathrm{[Fe/H]}<-1.5$, as shown by their results without subgrid redistribution model in post-processing.

The particle-based cosmological simulations of \citet{Haynes2019} show increasing $\mathrm{[\rp/Fe]}$ and $\mathrm{[\rp/\alpha]}$ with increasing $\mathrm{[Fe/H]}$ at subsolar metallicity for their neutron star merger model and a flat trend for magneto-rotational supernova model (which is similar to our fiducial rare core-collapse supernova model). These results also qualitatively agree with ours. However, our simulations do not exhibit the large scatter towards low r-process abundances at $-2<\mathrm{[Fe/H]}<0$ that is seen in their neutron star merger model. This could be due to differences in the chosen delay time distribution, the lack of mixing between gas particles, the limited resolution of their simulation, or a combination thereof. The inclusion of an on-the-fly subgrid metal diffusion model may further improve the agreement between our simulations.

Because our simulations are well-converged and the r-process enrichment was treated on-the-fly, with no imposed metallicity floor nor subgrid mixing prescription, we believe that our conclusions of a rising trend in $\mathrm{[\rp/Fe]}$ with metallicity for neutron star merger models is robust. Based on the fact that our r-process results vary little with resolution (see also Section~\ref{sec:res}), we expect that Lagrangian codes with realistic metal diffusion and on-the-fly r-process enrichment would also find an increasing trend of $\mathrm{[\rp/Fe]}$ with metallicity as well as somewhat reduced scatter compared to their previous results.

\section{Discussion and Conclusions} \label{sec:concl}

We studied the abundance of r-process elements in cosmological, magnetohydrodynamical simulations of Milky Way-mass galaxies using the moving mesh code \textsc{arepo} as part of the Auriga project. We implemented 10 models for r-process enrichment, 7 of which model neutron star mergers as the r-process production site and 3 of which assume that a fixed fraction of core-collapse supernovae are responsible for all of the r-process enrichment. These enrichment models are relatively simple, but show clearly what the results of different model assumptions are. The main conclusions from our present study are as follows.

\begin{enumerate}
  
\item Models with neutron star mergers as the only source of r-process elements result in the majority of stars being enriched with r-process elements, even at $\mathrm{[Fe/H]}<-3$, indicating that they could be important even in the early Universe when these stars were formed. However, the median r-process abundance ratio is below solar at low metallicity and increases with increasing metallicity. This is the case for all the model variations, which aim to bracket possible delay time distributions for neutron star mergers. 
  
\item Models that use rare core-collapse supernovae as the only r-process source show higher median r-process enrichment for metal-poor stars. The trend of $\mathrm{[\rp/Fe]}$ with metallicity is almost completely flat, although the rarest model shows a slight decrease towards the lowest metallicities. The scatter increases for models with fewer r-process events.
  
\item Metal-poor stars were formed at very high redshift, in the first few Gyr of the Universe. Outliers in $\mathrm{[\rp/Fe]}$ were formed at higher redshift than those following the median $\mathrm{[\rp/Fe]}$ trend. The majority of metal-poor stars were formed in external galaxies and accreted onto the main galaxy later on. Metal-poor stars that are outliers in $\mathrm{[\rp/Fe]}$ are more likely to have been accreted, but this residual trend is less strong than the one with redshift.

\item In general, Milky Way-mass galaxies with different formation histories have reasonably similar r-process abundance ratios at fixed metallicity, varying at most by a factor of 2 for $\mathrm{[Fe/H]}>-3$. At high metallicity and for models with relatively high rates of r-process producing events, the median r-process abundance ratios vary even less. 
  
\item Observations of the r-process element europium show increased scatter towards low metallicity, as do all of our models. The $\mathrm{[\rp/Fe]}$ trend with metallicity is approximately flat in observations for $\mathrm{[Fe/H]}<-1$. This seems to be more consistent with the rare core-collapse supernova models that closely follow the star formation rate of the galaxy than with neutron star merger models which show an increase of $\mathrm{[\rp/Fe]}$ with metallicity. However, more homogeneous observational data sets and potential improvements to the models may change this conclusion. 
  
\item Even though we found little variation between different Milky Way-mass galaxies, the r-process abundance ratios of dwarf galaxies can vary wildly for models with a relatively low rate of r-process producing events. If a dwarf galaxy in our simulations experiences an early r-process producing event in their evolution, their metal-poor stars are strongly r-process enhanced, as is also observed. As more stars form and more supernovae and r-process events occur and more material is mixed with the r-process enriched gas, the stellar abundances approach the median trend of the main Milky Way-mass galaxy. We studied the enrichment by a single neutron star merger in a simulated dwarf galaxy and found that the r-process elements mix with $\approx10^8$~M$_{\astrosun}$ of gas and $\approx10^2$~M$_{\astrosun}$ of iron within a few hundred~Myr, efficiently distributing r-process material over a large region. 

\end{enumerate}
  
All the neutron star merger models show a rising trend in $\mathrm{[\rp/Fe]}$ as a function of metallicity in our simulations and therefore seems to be robust. The simulations are well-converged, which lends additional credence to our results. However, the scatter in $\mathrm{[\rp/Fe]}$ increases slightly with improved resolution, especially the $2\sigma$ scatter. 

There are a few caveats to our study, however. The ISM model employed by our simulations does not attempt to capture the clumpy multi-phase structure of the Milky Way's ISM. The resulting model ISM is therefore smoother, possibly resulting in metals mixing too efficiently within the ISM (see also \citealt{Schonrich2019}), although it is also possible we underestimate mixing due to the absence of small-scale turbulence driven by local supernova feedback. Our simulations also do not include neutron star kicks, which could cause a larger fraction of neutron star mergers to occur outside the galaxy, in the low-metallicity halo. This would result in less efficient direct r-process enrichment of the ISM, but potentially higher r-process abundances in accreting gas. Whether or not adding such ingredients improves the match of our neutron star merger models to observations will be explored in future work.

Another possibility is that multiple types of sources are responsible for the r-process enrichment of the Universe, which could dominate at different epochs. A combination of neutron star mergers and rare core-collapse supernovae would likely be able to provide a reasonable match with observations of metal-poor stars. Future models could also, for example, include a metallicity dependence or non-uniform r-process yields depending on specific stellar parameters, which could change the scatter and the slope of the relation between $\mathrm{[\rp/Fe]}$ and $\mathrm{[Fe/H]}$. However, in this work we aimed to use simple models of single source types in order to study their resulting r-process enrichment. 

Any newly derived r-process yields for neutron star mergers or for a specific type of core-collapse supernova from either theory or observations, including potential dependence on metallicity, can also easily be incorporated into our cosmological framework. Ongoing and future spectroscopic surveys, such as Gaia-ESO, GALAH, WEAVE, and 4MOST, will obtain detailed abundance information of additional samples of metal-poor stars for both the Milky Way and nearby dwarf galaxies \citep[e.g.][]{Buder2018, Magrini2018, Christlieb2019}. Together with high-resolution simulations, these can be used to constrain both the early history of the Milky Way and its satellites as well as the production site of r-process elements.

\section*{Acknowledgements}

We would like to thank Jill Naiman and Andreas Bauswein for interesting discussions, Adrian Jenkins for providing the initial conditions for our simulations, and Tim Davis for comments on an earlier version of this manuscript. We would also like to thank the anonymous referee for helpful comments.
FvdV was supported by the Klaus Tschira Foundation and by the Deutsche Forschungsgemeinschaft through project SP 709/5-1.
FAG acknowledges financial support from CONICYT through the project FONDECYT Regular Nr. 1181264, and funding from the Max Planck Society through a Partner Group grant.
FM is supported by the program ``Rita Levi Montalcini'' of the Italian MIUR.
Our simulations were performed on computing resources provided by the Max Planck Computing and Data Facility in Garching.

\bibliographystyle{mnras}
\bibliography{rpauriga}

\bsp

\appendix

\section{Iron and magnesium} \label{sec:FeMg}

\normalsize

\begin{figure}
\center
\includegraphics[scale=.4]{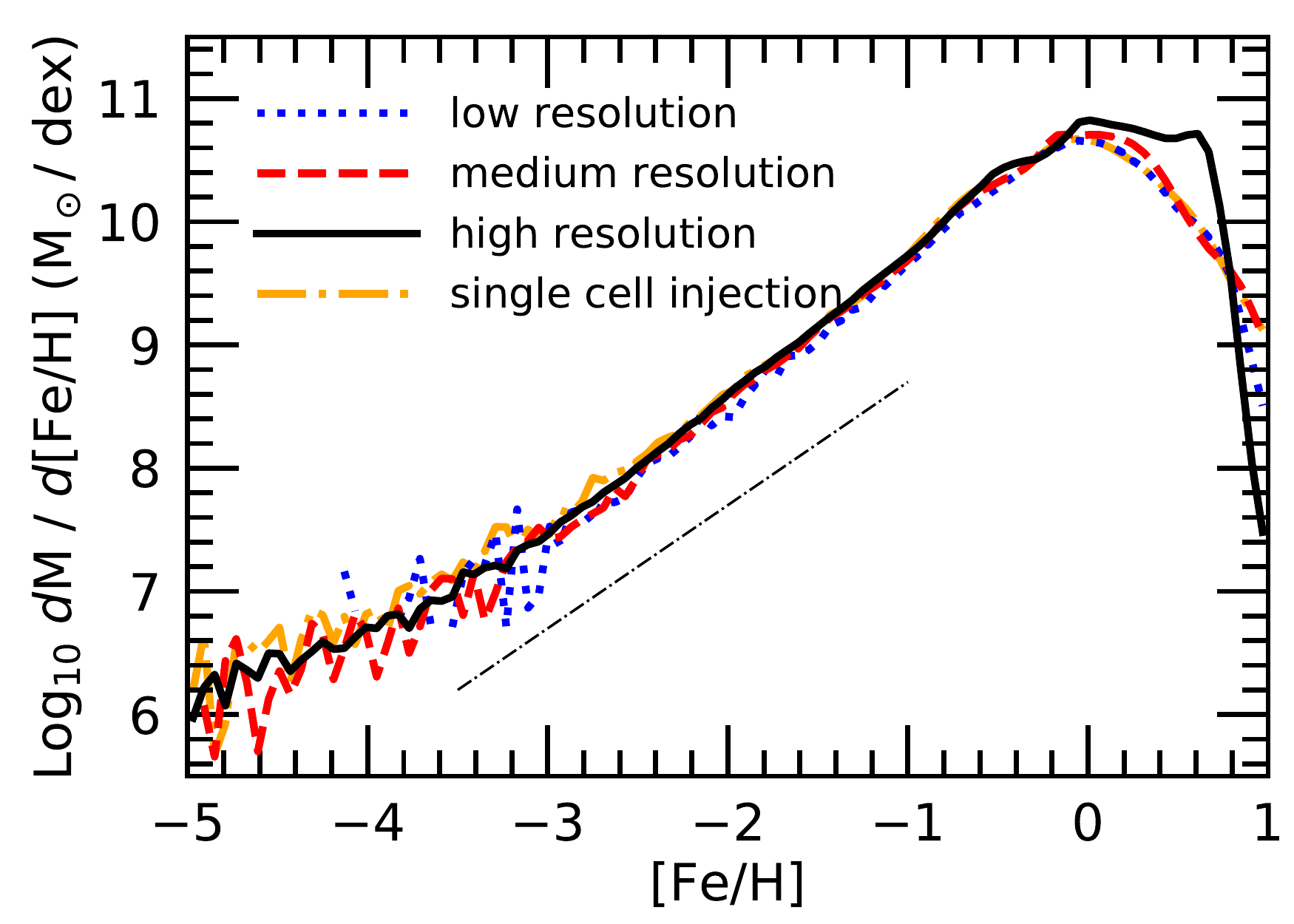} 
\caption {\label{fig:Fe} Logarithm of the metallicity distribution function (in solar masses per dex) of all stars within the virial radius at $z=0$ for three simulations spanning a factor 64 in mass resolution (linestyles are identical to those in Figure~\ref{fig:res}). The high resolution simulation (solid, black curve) has a somewhat higher star formation rate at low redshift than the lower resolution simulations and therefore forms slightly more metal-rich stars. The regime we study in this work, $\mathrm{[Fe/H]}<-1$, is converged. The metallicity distribution function is also independent of the initial metal injection region. The thin, dot-dashed line has a slope of unity, which is close to the slope of the distribution function. This means that there are about an order of magnitude more stars at $\mathrm{[Fe/H]}=-2$ than at $\mathrm{[Fe/H]}=-3$.} 
\end{figure}

The $\mathrm{[Fe/H]}$ distribution function for all the stars within the virial radius of our simulated galaxy is shown in Figure~\ref{fig:Fe} for high (solid, black curve), medium (dashed, red curve), and low (dotted, blue curve) resolution. The high-resolution simulation produces more stars with supersolar metallicities. This is because the galaxy has a higher star formation rate at low redshift than in the lower resolution simulations. Since we do not study high-metallicity stars in this work, we conclude that the metallicity distribution function is converged in the regime we focus on.

The dot-dashed, orange curve shows the result for a medium resolution simulation where iron (and other metals) were injected only in the single host cell of the star that produced them, rather than in 64~cells (i.e.\ the host cell plus 63~neighbouring cells). Because of the exchange of mass and metals between cells, the metals mix with lower metallicity gas, and the resulting metallicity distribution function is insensitive to reducing the initial enrichment region. 

The thin, dot-dashed line indicates a slope of unity, which is closely followed by the metallicity distribution function for subsolar metallicities. The distribution of metal-poor stars in our simulations, for which the number of stars decrease by an order of magnitude for every dex decrease in metallicity, is in agreement with some observational samples \citep{Yong2013, Li2018}. However, other samples show a cut-off at low metallicity \citep{Schorck2009}, potentially due to the way the sample was selected or differences in analysis of the data \citep{Yong2013}. Larger samples of stars, that span an even wider range in metallicity, will be necessary in order to confirm the match with the metallicity distribution function obtained from our simulations.

\begin{figure}
\center
\includegraphics[scale=.4]{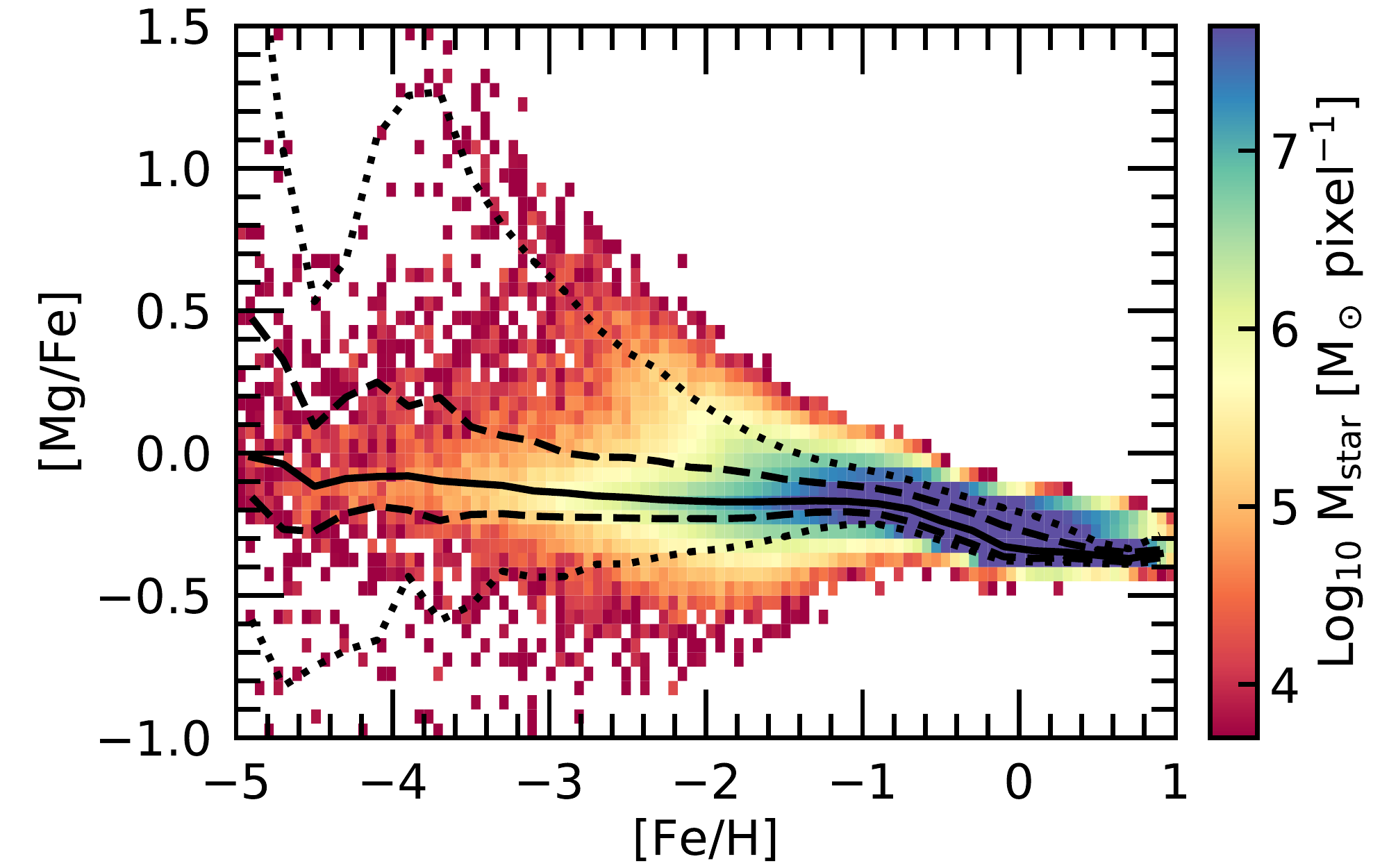} 
\caption {\label{fig:Mg} Magnesium abundance ratio: $\mathrm{[Mg/Fe]}$ against $\mathrm{[Fe/H]}$. The colour coding and curves have the same meaning as in Figure \ref{fig:NS}. The scatter in the magnesium abundance ratio is much smaller than for $\mathrm{[\rp/Fe]}$, indicating that the magnesium production site is much more common. The normalization is about 0.4~dex too low as compared to observations, which is a known problem for the yields of \citet{Portinari1998} used in our simulations, but this does not affect our conclusions. In the main body of the paper, we correct for this discrepancy by multiplying the magnesium yields by a factor~2.5 in post-processing. The decrease at high metallicity is not strong enough, likely due to the incorrect balance of iron produced by Type Ia supernovae and core-collapse supernovae. We therefore focus on the regime where core-collapse supernovae dominate, $\mathrm{[Fe/H]}<-1$.} 
\end{figure}

Figure~\ref{fig:Mg} shows the abundance ratio of the $\alpha$ element magnesium to iron as a function of iron metallicity. The $1\sigma$ scatter is small accross the full range in metallicity and the median $\mathrm{[Mg/Fe]}$ is approximately constant for $\mathrm{[Fe/H]}<-1$, in agreement with observations of the Milky Way \citep[e.g.][]{Cayrel2004, Arnone2005}. The normalization of $\mathrm{[Mg/Fe]}$ is, however, lower than in observations, by about 0.4~dex. This is caused by a well-known underproduction of magnesium in the adopted yields \citep{Portinari1998}. Because magnesium is not a strong coolant in our simulations, this is not expected to affect the dynamics of our simulations and can therefore be corrected in post-processing. In the main text, we show results for which the magnesium yield was multiplied by a factor~2.5. The values of $\mathrm{[Mg/H]}$ in Figures~\ref{fig:NS} and~\ref{fig:SN} were therefore increased by 0.4~dex in post-processing. We use $\mathrm{[\rp/Mg]}$ (after correcting the magnesium yields) to renormalize the r-process abundances throughout this work by setting $\mathrm{[\rp/Mg]}$ to zero at $\mathrm{[Mg/H]}=0$ and using the same r-process yiels for $\mathrm{[\rp/Fe]}$. This results in different r-process (or europium) yields for each r-process enrichment model, as listed in Tables~\ref{tab:NSmodels} and~\ref{tab:SNmodels}. In this work, we study the trends of r-process abundance ratios with metallicity and not the normalization, owing to the uncertain r-process yields. The renormalization in post-processing therefore does not affect our conclusions.

At high metallicity, observations show a stronger decrease of $\mathrm{[Mg/Fe]}$ with increasing $\mathrm{[Fe/H]}$ than our simulations. As discussed in Section~\ref{sec:results}, this could be either due to the underproduction of iron by Type Ia supernovae or the overproduction of iron by core-collapse supernovae. Because of this, we focus our analysis of r-process elements entirely on the regime where core-collapse supernovae are known to dominate, i.e.\ at $\mathrm{[Fe/H]}<-1$.

\label{lastpage}

\end{document}